\documentclass{pasj01}
\Received{$\langle$reception date$\rangle$}
\Accepted{$\langle$acception date$\rangle$}
\Published{$\langle$publication date$\rangle$}
\usepackage{color} 
\usepackage{graphicx} 
\usepackage{multirow}
\usepackage{multicol}
\usepackage{changepage}

\begin{document}

\title{{High-mass star formation in the Large Magellanic Cloud triggered by colliding H{\sc i} flows}}
\author{Kisetsu Tsuge\altaffilmark{1,2,3,4}, Hidetoshi Sano\altaffilmark{3}, Kengo Tachihara\altaffilmark{5}, Kenji Bekki\altaffilmark{6}, Kazuki Tokuda\altaffilmark{7,8}, Tsuyoshi Inoue\altaffilmark{9}, Norikazu Mizuno\altaffilmark{7}, Akiko Kawamura\altaffilmark{7}, Toshikazu Onishi\altaffilmark{10}, Yasuo Fukui\altaffilmark{5}}%
\altaffiltext{1}{Department of Physics, Graduate School of Science, the University of Tokyo, 7-3-1 Hongo, Bunkyo-ku, Tokyo 113-0033, Japan}
\altaffiltext{2}{Institute for Advanced Study, Gifu University, 1-1 Yanagido, Gifu 501-1193, Japan}
\altaffiltext{3}{Faculty of Engineering, Gifu University, 1-1 Yanagido, Gifu 501-1193, Japan}
\altaffiltext{4}{Institute for Advanced Research, Nagoya University, Furo-cho, Chikusa-ku, Nagoya 464-8601, Japan}
\altaffiltext{5}{Department of Physics, Nagoya University, Furo-cho, Chikusa-ku, Nagoya 464-8601, Japan}
\altaffiltext{6}{ICRAR, M468, The University of Western Australia, 35 Stirling Highway, Crawley Western Australia 6009, Australia}
\altaffiltext{7}{National Astronomical Observatory of Japan, Mitaka, Tokyo 181-8588, Japan}
\altaffiltext{8}{Department of Earth and Planetary Sciences, Faculty of Science, Kyushu University, Nishi-ku, Fukuoka 819-0395, Japan}
\altaffiltext{9}{Department of Physics,Faculty of Science and Engineering,Konan University, 8-9-1 Okamoto, Higashinada-ku Kobe 658-8501, Japan}
\altaffiltext{10}{Department of Physics, Graduate School of Science, Osaka Metropolitan University, 1-1 Gakuen-cho, Naka-ku, Sakai, Osaka 599-8531,Japan}

\email{tsuge.kisetsu.i2@f.gifu-u.ac.jp}


\maketitle

\begin{abstract}
The galactic tidal interaction is a possible mechanism to trigger the active star formation in galaxies. The recent analyses using the H{\sc i} data in the Large Magellanic Cloud (LMC) proposed that the tidally driven H{\sc i} flow, the L-component, is colliding with the LMC disk, the D-component, and is triggering high-mass star formation toward the active star-forming regions R136 and N44. In order to explore the role of the collision over the entire LMC disk, we investigated the I-component, the collision-compressed gas between the L- and D-components, over the LMC disk, and found that 74 \% of the O/WR stars are located toward the I-component, suggesting their formation in the colliding gas. {We compared four star-forming regions (R136, N44, N11, {N77-N79-N83 complex}). We found a positive correlation between the number of high-mass stars and the compressed gas pressure generated by collisions, suggesting that the pressure may be a key parameter in star formation.}
\end{abstract}

\section{Introduction} \label{sec:intro}
\subsection{Active star formation induced by the galactic tidal interaction}
Starburst is one of the {critical} processes in the galaxy{'s} evolution and the star formation history of the {U}niverse. Many early studies suggested that tidal perturbations from nearby companions \citep{1988A&A...203..259N,1990PASJ...42..505F}, galactic interactions {including} mergers (e.g., \cite{1988ApJ...325...74S,2008MNRAS.388L..10B}), or cold gas accretion from the intergalactic medium (e.g.,\cite{1987ApJ...322L..59S}) are possible mechanisms triggering the burst of star formation. \citet{1998Natur.395..859G} show{ed} that galaxy mergers have excess infrared luminosity compared with isolated galaxies, lending support {to} the active star formation triggered by galaxy interactions. {Subsequently}, \citet{2014A&A...566A..71L} investigated 18 starburst dwarf galaxies using H{\sc i} data. They found that starburst dwarf galaxies have more asymmetric H{\sc i} morphologies than typical dwarf irregulars, and $\sim$80 \% of the starburst dwarf galaxies are interacting galaxies with at least one potential companion within 200 kpc. Thus, these previous works suggest that some external mechanism induced by galactic interaction triggers the starburst. In addition, numerical simulations of galaxy interactions show that interactions/mergers between gas-rich dwarfs formed irregular blue compact dwarfs (BCDs) , including IZw~18, which hosts starburst \citep{2008MNRAS.388L..10B}. BCDs {have} low metallicity (0.1$\geq$$Z$/$Z_{\rm \odot}$$\geq$0.02) and {are} similar to the environment where the first stars formed in the early {U}niverse \citep{1972ApJ...173...25S}. Therefore, {elucidation of} the triggering mechanism of active star formation in dwarf galaxies has the potential to promote understanding of the origin of starbursts in the early {U}niverse. Most of the interacting galaxies are, however, distant, and it is difficult to resolve individual clouds and investigate their physical properties in detail{, except for recent ALMA studies toward {the} nearby interacting system, such as the Antennae Galaxies (e.g., \cite{2014ApJ...795..156W,2021PASJ...73S..35T,2021PASJ...73..417T})}.

The present study focuses on the Large Magellanic Cloud (LMC). The LMC is one of the nearest interacting dwarf galaxies (distance 50$\pm$1.3 kpc;\cite{2013Natur.495...76P}) and is almost face-on with an inclination of $\sim$20--30 deg. (e.g., \cite{2010A&A...520A..24S}). 
{Thus, the LMC is an optimal laboratory for investigating the active star formation mechanisms across cosmic history, covering a wide spatial-dynamic range from a galactic scale (kpc scale) down to a cluster scale (10--15 pc).} The mean metallicity of the LMC is approximately half of the solar metallicity (0.3--0.5 $Z_{\rm \odot}$; \cite{1992ApJ...384..508R, 1997macl.book.....W}), which is close to the mean metallicity of the interstellar medium during the time of peak star formation (redshift z$\sim$1.5;  \cite{1999ApJ...522..604P}). {The} LMC is an optimal laboratory {for} investigating the active star formation mechanism{s} through a wide {spatial-dynamic} range from a galactic scale (kpc scale) down to a {stellar} cluster scale (10--15 pc).

There are many active star-forming regions over the LMC, which have been intensively studied ({N77-N79-N83 complex}: e.g., \cite{2017NatAs...1..784O}; Nayak et al. 2019, N159: Chen et al. 2010; Saigo et al. 2017; Fukui et al. 2019; Tokuda et al. 2019, N11: e.g., Walborn \& Parker 1992; Celis Pena et al. 2019, N44; e.g., Chu et al. 1993; Chen et al. 2009, N206: e.g., Romita et al. 2010, N51: e.g., Lucke \& Hodge 1970; Chu et al. 2005, N105:e.g., Epchtein et al. 1984; Oliveira et al. 2006, N113: e.g., Brooks \& Whiteoak 1997, N120: e.g., Lucke \ Hodge 1970; N144; e.g., Lortet \& Testor 1988 etc. cataloged by Henize 1956 {listed in ascending R.A. order}). Above all, recent works by Fukui et al. 2017 and Tsuge et al. 2019 revealed evidence for the triggered formation of the massive {young} cluster R136, the H{\sc ii} region N44 and some of the other high-mass stars in the LMC. These authors suggested that the tidal interactions induced colliding H{\sc i} gas flows in the LMC.

\subsection{The interacting two velocity components}
{Many observational studies support the LMC--SMC interaction from the stellar proper motion {of the Magellanic Bridge}. {The Magellanic Bridge is a structure of H{\sc i} gas that extends like a bridge between the LMC and the SMC, believed to have formed due to the gravitational interaction between the LMC and SMC (Murai \& Fujimoto, 1980; Gardiner et al., 1994).} The latest observations of stellar proper motion using GAIA suggested a close encounter and collision between the SMC and LMC $\sim$0.2 Gyr ago (e.g., \cite{2019ApJ...874...78Z,2013ApJ...764..161K}). {The authors of Kallivayalil et al. (2013)} revealed the proper motions of stars in the Magellanic Bridge and found that the stars moved from the SMC to the LMC. They compared their results with a numerical model of the Magellanic Bridge formation and {discovered} that the observations agree with a model in which the SMC-LMC collision occurs with an impact parameter of less than a few kpc (e.g., \cite{2018ApJ...867L...8O}).}

{The star formation history estimated from photometry of stellar populations also supports the LMC--SMC interaction as shown by numerical simulations (e.g., \cite{2009ApJ...705.1260N,2020A&A...641A.134S,2005MNRAS.358.1215P,2005AJ....130.1083M,2009AJ....138.1243H,2018ApJ...866...90C,2022ApJ...927..153C,2018ApJ...864...55Z}). {Especially, \citet{2009ApJ...705.1260N} and \citet{2009AJ....138.1243H} support the recent interaction within 1 Gyr}. {These simulation models need to be confronted with observations not only of the stars but also of the distribution and kinematics of the gas that forms stars. This contrasts with previous simulation studies, which are based on comparisons with observations of stars only.} \citet{2022ApJ...927..153C} also suggests that the best model is one in which the latest collision occurred within ~0.25 Gyr with the collision parameter of 10 kpc. We should note that the kinematics and distribution of the stars are the product in a longer time scale than the gas. {Fukui et al. (2018) estimated the ionization velocity by O-stars to be 5 km/s by comparing observational results of molecular clouds associated with massive star clusters of various ages. Based on this velocity, it is estimated that molecular clouds within 50 pc of the star cluster will be ionized after 10 million years, making verifying the gas involved in star formation challenging.} It is, therefore, essential to look at the associated gas component, which reflects events in a short time scale of 10 Myr, in order to elucidate detailed dynamics of the interaction and its relationship to recent star formation}

\begin{figure*}[htbp]
\begin{center}
\includegraphics[width=\linewidth]{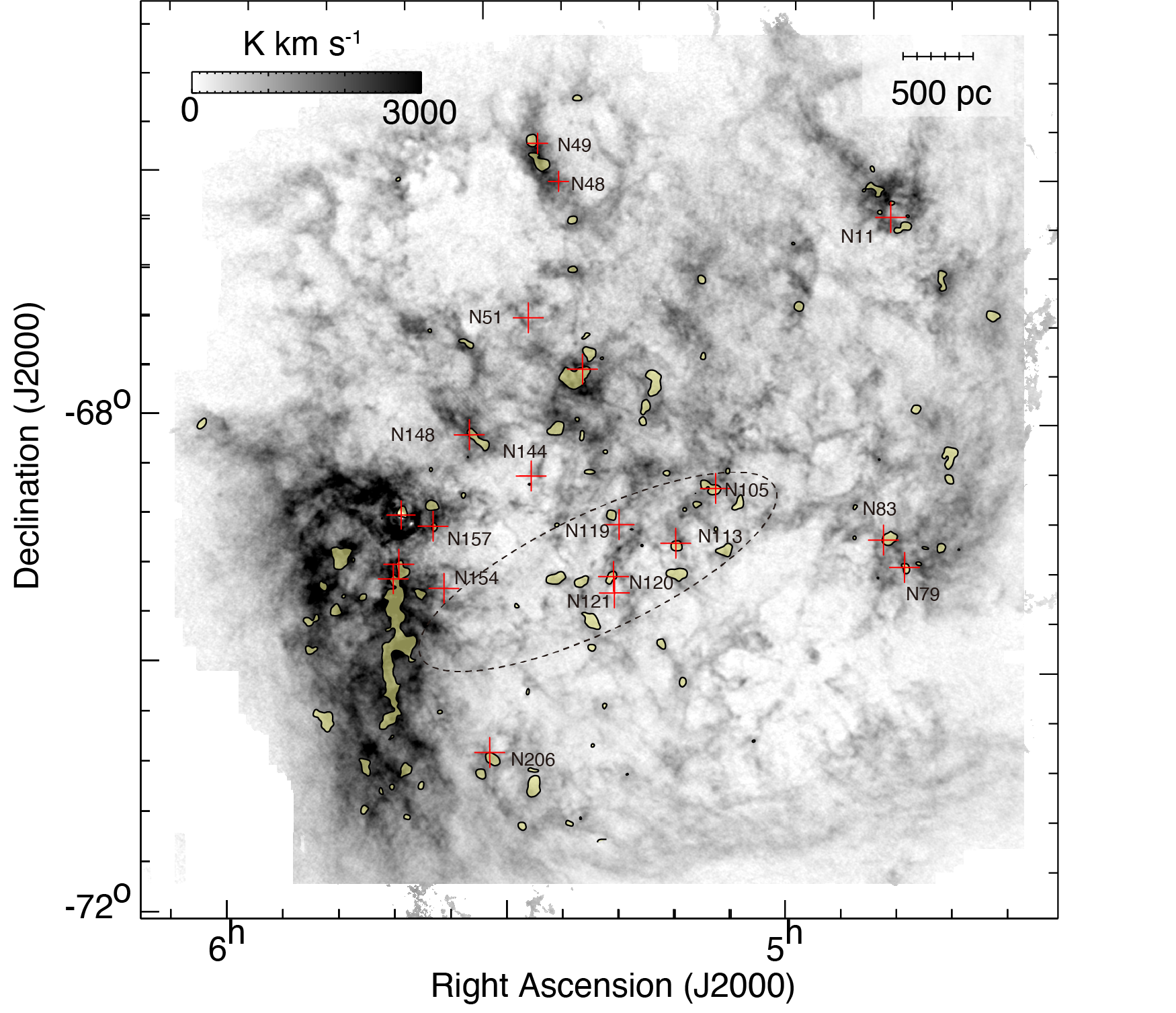}
\end{center}
\caption{The H{\sc i} integrated intensity maps with the integration velocity range of $V_{\rm offset}$=$-$101.1--29.9 km s$^{-1}$.  The yellow shaded areas are the regions where $^{12}$CO($J$=1--0) intensity obtained with NANTEN is greater than 1.5 $\sigma$ (1.48 K km s$^{-1}$) with the same velocity range. Plus signs show the positions of luminous H{\sc ii} regions (S$_{H\alpha}$ $>$ 1$\times$10$^{4}$ erg cm$^{-2}$ s$^{-1}$ sr$^{-1}$; Ambrocio-Cruz et al. 2016). {The dashed line shows the optical boundary of the stellar bar.}}  
\label{fig1}
\end{figure*}%

{\citet{1992A&A...263...41L} identified two velocity H{\sc i} components based on H{\sc i} observations at 15 $\arcmin$ resolution {(corresponding to $\sim$225 pc at the distance of the LMC)} and named the two as the L-component and the D-component, where the L-component has smaller velocity by 50 km s$^{-1}$ than the D-component. Subsequently, H{\sc i} observations at a higher spatial resolution of 1 $\arcmin$ were conducted by \citet{2003ApJS..148..473K}, and H{\sc i} shells and holes were investigated in detail over the LMC{,  as shown in Figure 1. The presence of multiple components results in an asymmetric intensity distribution of the H{\sc i} gas shown in Figure 1}. Signs for the L- and D-components were also recognized in the high-resolution data. {In contrast,} the} {previous studies including \citet{2003MNRAS.339...87S} and \citet{2015A&A...573A.136I} did not investigate the physical properties of the two components.}

{Most recently, the two components were reviewed in the light of the colliding H{\sc i} flows which triggered the formation of {Young Massive Clusters (YMCs)} by} \citet{2017PASJ...69L...5F} (hereafter Paper I) and by \citet{2019ApJ...871...44T} (hereafter Paper II). {YMCs are astronomical objects characterized by total cluster masses typically exceeding 10$^4$ $M_{\odot}$ and containing many high-mass stars within a radius of 1 pc (Portegies Zwart et al., 2010)} {These works revealed that the L- and D-components are colliding toward two outstanding clusters/H{\sc ii} regions, R136 {in the H{\sc i} Ridge} and N44 {in the northern part}. {They} suggested that the collisional trigger likely forms both. }

Conventionally, the rotation curve of the LMC was obtained from the H{\sc i} data, where the L- and D-components are mixed up. In their study, Paper II developed a method to decompose the L- and D-components by subtracting the galactic rotation, which laid a foundation for a detailed kinematical study of H{\sc i} in the LMC. {This rotation curve is consistent with the latest results presented {by} Oh et al. (2022)}. They thereby confirmed the original suggestion by \cite{1992A&A...263...41L} on the two components, and identified observational signatures of the collision between the two components as follows; i. the \underline{complementary spatial distribution} between the L- and D- components and ii. the \underline{intermediate velocity component} (hereafter the I-component) connect{s} the two components in velocity space, which support{s} the collisional interaction of the L- and D-components. 
By using the Atacama Large Millimeter/submillimeter Array (ALMA), \citet{2019ApJ...886...14F,2019ApJ...886...15T} {found massive filamentary clouds {toward the N159 region in the H{\sc i} Ridge,} holding high-mass star formation, which shows cloud collision signatures at a few pc in the CO clouds}.

Based on these signatures, Papers I and II, \citet{2019ApJ...886...14F}, and \citet{2019ApJ...886...15T} argued that the collision between the L- and D-components worked as a formation mechanism of $\sim$400 high-mass stars including R136, N159, and N44. {This scenario is supported by the} numerical simulations {which} elaborate the observational characteristics of a cloud-cloud collision (e.g., \cite{1992PASJ...44..203H,2010MNRAS.405.1431A,2014ApJ...792...63T,2018PASJ...70S..58T,2018PASJ...70S..54S,2018PASJ...70S..53I,2018PASJ...70S..59K,2021PASJ...73S.385S,2021ApJ...908....2M,2018ApJ...859..166F,2021PASJ...73S.405F}; see also for a review \cite{2021PASJ...73S...1F}).

{The origin of the L-component is most likely the tidal interaction between the LMC and the SMC at their close encounter, which occurred $\sim$0.2 Gyr ago. The gas stripped from both the LMC and the SMC by the tidal force is expected to fall down currently to the LMC disk, which causes the L- and D-components. This scenario is supported by the detailed numerical simulations of the tidal interaction by \cite{2007PASA...24...21B} (see also \cite{2007MNRAS.381L..16B,2005MNRAS.356..680B,2014MNRAS.443..522Y}). Papers I and II developed the discussion along the tidal interaction and argued for the gas injection from the SMC into the LMC based on the low dust-to-gas ratio. 
{Paper I and Paper II estimated the metal amount by a comparison of dust optical depth at 353 GHz ($\tau$353) measured by $Planck/IRAS$ telescopes (Planck Collaboration 2014) and the intensity of H{\sc i} ($W$(H{\sc i})). The authors found a factor of two differences in the dust-to-gas ratio between the L-component and the D-component. That difference corresponds to the difference in metallicity if we assume that the dust-to-gas ratio is constant in the LMC. The L-component is estimated to have about half the metallicity of the D-component due to gas inflow from the SMC (Paper I).}

\begin{figure*}[htbp]
\begin{center}
\includegraphics[width=\linewidth]{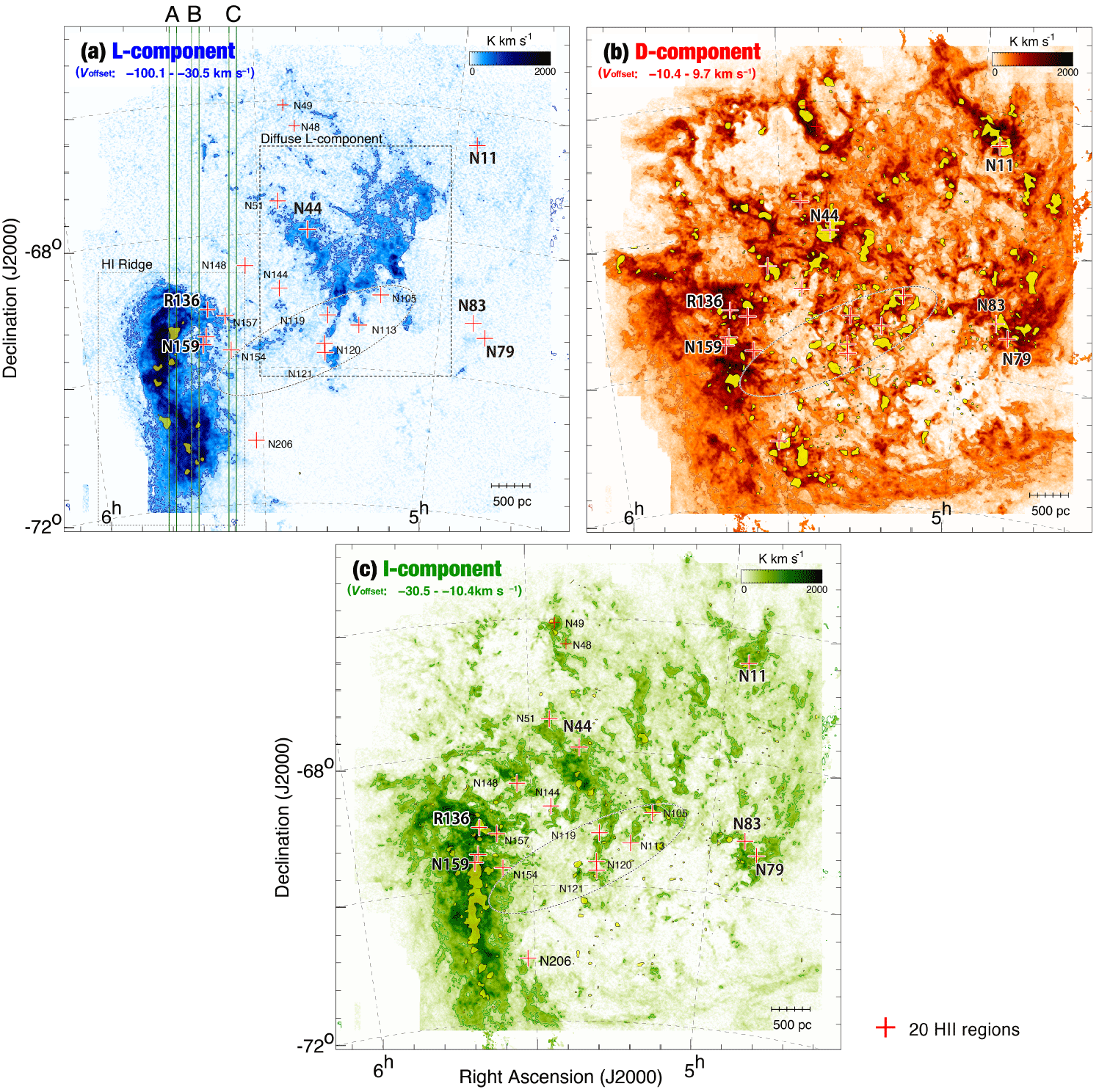}
\end{center}
\caption{The H{\sc i} integrated intensity maps of (a) the L-component, (b) the D-component, and (c) the I-component. The integration velocity range is  $V_{\rm offset}$=$-$101.1--$-$30.5 km s$^{-1}$ for the L-component, $V_{\rm offset}$=$-$10.4--9.7 km s$^{-1}$ for the D-component, and $V_{\rm offset}$=$-$30.5--$-$10.4 km s$^{-1}$ for the I-component. The lowest contour level and intervals are 192 K km s$^{-1}$ (15 $\sigma$) and 769 K km s$^{-1}$ (60 $\sigma$) for (a) and 279 K km s$^{-1}$ (40 $\sigma$) and 419 K km s$^{-1}$ (60 $\sigma$) for (b) and (c). The {yellow} shaded areas in three panels are the regions with $^{12}$CO($J$=1--0) intensity greater than 1.5 $\sigma$ with the same velocity range. Plus signs show the positions of luminous H{\sc ii} regions (S$_{H\alpha}$ $>$ 1$\times$10$^{4}$ erg cm$^{-2}$ s$^{-1}$ sr$^{-1}$; Ambrocio-Cruz et al. 2016). The green lines in (a) show the integration ranges in R.A. in Figure \ref{fig21}.}  
\label{fig2}
\end{figure*}

\subsection{The I-component: a possibl{e} tracer of the colliding H{\sc i} flows {and triggered star formation}}\label{sec:1.3}
We defined the I-component, whose velocity is intermediate between the L- and D-components in Paper II. We interpret that the I-component is {the mixture of} {the} decelerated L-component and the D-component in the collisional interaction. According to the picture, molecular gas is formed by the density increase in the compressed interface layer, as shown in the synthetic observations by \citet{2018ApJ...860...33F} of the colliding H{\sc i} flows which were numerically simulated by \citet{2012ApJ...759...35I}. It is likely that the molecular gas leads to the formation of massive stellar clusters \citep{2021ApJ...908....2M}. Such a high-density part of the I-component is found in the CO distribution toward R136, N159, and N44 (Figures \ref{fig8new} and \ref{fig8}a and Paper II), suggesting the possibility that high-mass star formation is triggered by the colliding H{\sc i} flows over the LMC. {Tsuge et al. (2021) found that the number density of colliding gas and collision velocity are important parameters, and there is a positive correlation between the collisional compression pressure calculated from the density and velocity and the mass of the stellar cluster.} {The present study }aims to explore the role of the collisional interaction of the H{\sc i} gas and its relationship with high-mass star formation over the whole LMC. 

{T}o achieve this goal, we analy{ze} H{\sc i} data {comprehensively}. We summarize the quantitive comparison of spatial distributions of the I-component and high-mass stars in Section \ref{sec:3.3}. A detailed investigation of the observational signatures of H{\sc i} collisions in {the} N11 and {N77-N79-N83 complex} in Section \ref{sec:4.1.3} and \ref{sec:4.1.4}, respectively. N11 is located at the edge of supergiant shell 1 (Dawson et al. 2013) and is the oldest H{\sc ii} region among these regions. {N77-N79-N83 complex} is located at the origin of the western tidal arm and, {N79} is noted as a feature rival to 30Dor (Ochsendorf 2017). We cover a wide range of locations and surrounding environments {(density of colliding gas, collision velocities, collisional compression pressures, and metallicity)} in the LMC, as shown in Figure 1.} { These differences can influence important physical quantities related to star formation and molecular cloud formation, such as gas mass accretion rates and cooling efficiency.} 

\begin{figure*}[htbp]
\begin{center}
\includegraphics[width=12cm]{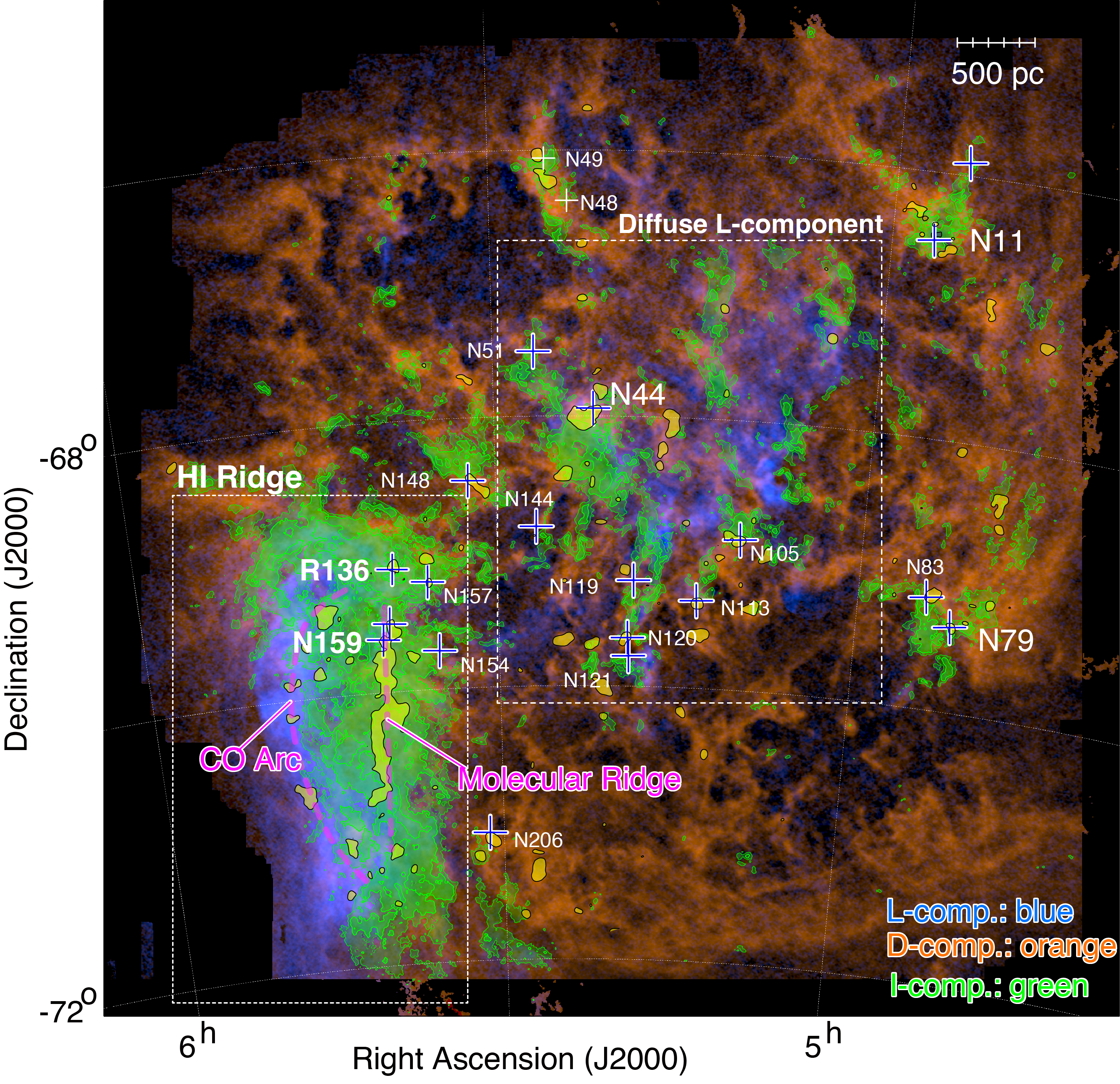}
\end{center}
\caption{H{\sc i} intensity map of the I-component by contours superposed on the L- and D-components image. The velocity ranges of three components are the same as in Figure 1. The contour levels are 300, 500 and 1000 K km s$^{-1}$. Plus signs show the positions of luminous H{\sc ii} regions (S$_{H\alpha}$ $>$ 1$\times$10$^{4}$ erg cm$^{-2}$ s$^{-1}$ sr$^{-1}$; \cite{2016MNRAS.457.2048A}). {Molecular clouds of N48 and N49, which are located between two SGSs, LMC 4 and LMC 5 are also shown by crosses}. The shaded areas in yellow are the the regions where $^{12}$CO($J$=1--0) intensity obtained with NANTEN is greater than 1.5 $\sigma$ (1.48 K km s$^{-1}$) with the same velocity range. {The magenta dashed lines shows the Molecular Ridge and CO-Arc region.}}  
\label{fig3}
\end{figure*}%

\section{Datasets} \label{sec:data}
{We used the original angular resolution to compare the spatial distributions, whereas we smoothed the data and matched the resolution of CO and H{\sc i} to calculate physical parameters.}
\subsection{HI}\label{sec:2.1}
We used archival data of H{\sc i} 21 cm line emission of the whole LMC obtained with Australia Telescope Compact Array (ATCA) and Parkes telescope \citep{2003ApJS..148..473K}. {They combined the H{\sc i} data obtained by ATCA (Kim et al. 1998) with those obtained with the Parkes multi beam receiver with a resolution of 14$\arcmin$--16$\arcmin$\citep{1997PASA...14..111S}.} The angular resolution of the combined H{\sc i} data is 60$\arcsec$ (corresponding to $\sim$15 pc at the distance of the LMC). The rms noise level is 2.4 K {at} a velocity resolution of 1.649 km s$^{-1}$. More detailed descriptions of the observations are given by \citet{2003ApJS..148..473K}.

\subsection{CO}\label{sec:2.2}
$^{12}$CO($J$=1--0) data {obtained with the} NANTEN 4 m telescope \citep{1999PASJ...51..745F, 2008ApJS..178...56F, 2001PASJ...53..971M} are used for a large-scale analysis. {These} observations cover a 6$^{\circ}$$\times$6$^{\circ}$ area, including the whole optical extent of the LMC, {are} suitable for the comparison with {the} kpc scale H{\sc i} dynamics.  The half-power beam width {was} 2$\farcm$6 {(corresponding to $\sim$40 pc at the distance of the LMC)} with {a regular} grid spacing of 2$\farcm$0, and {a} velocity resolution {was} 0.65 km s$^{-1}$.

We also used $^{12}$CO($J$=1--0) data of the Magellanic Mopra Assessment (MAGMA; \cite{2011ApJS..197...16W}) for a small-scale analysis in each star-forming region. The angular resolution {was} 45$\arcsec$ (corresponding to 11 pc at the distance of the LMC),{ and} {the} velocity resolution {was} 0.526 km s$^{-1}$. The MAGMA survey does not cover the whole LMC, and the observed area is limited {to} the individual CO clouds detected by {the} NANTEN survey.

\subsection{H$\alpha$}\label{sec:2.3}
We used the H$\alpha$ data obtained by the Magellanic Cloud Emission-Line Survey (MCELS; \cite{1999IAUS..190...28S}). The dataset was obtained with a 2048$\times$2048{-pixel} CCD camera on the Curtis Schmidt Telescope at Cerro Tololo Inter-American Observatory. The angular resolution {was} $\sim$3$\arcsec$--4$\arcsec$ (correspond{ing} to $\sim$0.75--1.0 pc at a distance of the LMC). We also use the archival data of H$\alpha$ provided by the Southern H-Alpha Sky Survey Atlas (SHASSA; \cite{2001PASP..113.1326G}) to define the region where UV radiation is locally enhanced by star formation.

\section{Observational Results}\label{sec:results}

\subsection{Spatial- and velocity {distributions} of HI gas at a kpc scale}\label{sec:3.1}
{The method of derivation of the three components is summarized below. The L- and D-components were decomposed over the whole LMC for the first time at a high angular resolution of 1$\arcmin$ {(corresponding to $\sim$15 pc at the distance of the LMC)} in Papers I and II. The rotation curve of the D-component was then derived in Paper II, which corresponds to the LMC disk. {The rotation curve differs from the previous works \citep{2003ApJS..148..473K} which} treated all the H{\sc i} components as the LMC disk. {The present rotation curve includes only the D-component.} We then defined $V_{\rm offset}$ as a relative velocity from the rotational velocity of the D-component ($V_{\rm offset}$ = $V_{\rm LSR}$$-$$V_{\rm D}$; $V_{\rm D}$ = the projected rotation velocity of the D-component). The integration ranges are $V_{\rm offset}$: $-$100.1--$-$30.5 km s$^{-1}$ for the L-component; and $V_{\rm offset}$: $-$30.5 -- $-$10.4 km s$^{-1}$ for the I-component; $V_{\rm offset}$: $-$10.4--9.7 km s$^{-1}$ for the D-component as explained in Paper II. We also made a histogram of H{\sc i} gas toward the northern part of the H{\sc i} Ridge region, as shown in Figure \ref{fig16} of Appendix 1. This histogram illustrates that the velocities $-$30 km s$^{-1}$ and $-$10 km s$^{-1}$ well correspond to the boundaries between the three components.}

{Figure 2 shows overlays of the distribution of the three H{\sc i} components with the major 20 star-forming regions ($S_{\rm H\alpha}$ $>$ 1$\times$10$^4$ erg cm$^{-2}$ s$^{-1}$ sr$^{-1}$; Ambrocio-Cruz et al. 2016) over the whole LMC. }
We present the distributions of {the three components in} Figures \ref{fig2}(a), \ref{fig2}(b), and \ref{fig2}(c), {which show the spatial} distributions of the L-, D-, and I-components, respectively. Figure \ref{fig3} is an overlay of {high integrated intensity areas ($W$(H{\sc i})$>$300 K km s$^{-1}$)} of the three {distributions}. The distributions of the L-, I-, and D-components are significantly different, while these three components partially show similar distribution. All three components, L, I, and D, are concentrated in the H{\sc i} Ridge region. While the D-component extends across the entire galaxy, the L-component is concentrated in the southeastern H{\sc i} Ridge and the northwestern Diffuse L-component directions. The I-component is concentrated toward regions where the intensity of the D-component is outstanding. The spatial distribution of active H{\sc ii} regions which have surface luminosity $S_{\rm H\alpha}$ greater than 1$\times$10$^4$ erg cm$^{-2}$ sr$^{-1}$ \citep{2016MNRAS.457.2048A} resembles the strong I-component. 



\subsection{{Detailed properties of the three HI components}}\label{sec:3.2new}
\underline{The L-component} is composed of two {kpc-scale} extended {features}. One is the H{\sc i} Ridge region located in the southeast region{, which includes} two major elongated CO {components}, i.e., the Molecular Ridge and {the} CO-Arc (\cite{1999PASJ...51..745F,2001PASJ...53..971M}, {magenta dashed lines of Fig{ure} 3}). The other is the Diffuse component extending {toward} the northwest as {shown in {the} dashed box of Figure \ref{fig2}a} (hereafter {the} Diffuse L-component {as in Paper II}). \underline{The I-component} {is distributed along the western rim of H{\sc i} Ridge's L-component and the Diffuse L-component's periphery, as shown} in Figure \ref{fig3}. The I-component is also located toward the southern end of {the} western arm \citep{2003MNRAS.339...87S}, including {N77-N79-N83 complex} and N11, {where we find only weak signs of the L-component}. {The I-component shows good correspondence with the Molecular Ridge, while it has little {resemblance} with the CO-Arc. Finally, \underline{the D-component} is distributed over the whole LMC.}

{H{\sc i} and H$_{2}$ masses} {($M$(H{\sc i}) and $M$(H$_{2}$))} of the L-, D-, and I-components are summarized for the whole LMC and the H{\sc i} Ridge region, respectively, in Table \ref{tab:mass}. We calculate{d} {the} mass of the H{\sc i} gas in the assumption that H{\sc i} emission is optically thin as follows, 
\begin{equation}
{M{\rm (HI)} = m_{p} \Omega D^2 \Sigma_{i} N_{i}({\rm HI})},
\end{equation}
{where $m_p$ is the mass of hydrogen. $D$ is the distance to the source in cm, equal to 50 kpc, $\Omega$ is the solid angle subtended by a unit grid spacing of a square pixel, and $N_i$(HI) is the atomic hydrogen column density for each pixel in cm$^{-2}$.}

\begin{equation}
{N({\rm H{\sc I})} = 1.8224\times10^{18}\int \Delta T_{\rm b}  \it dv \ \rm{[cm^{-2}] },}
\end{equation}
where $T_{\rm b}$ is the observed H{\sc i} brightness temperature (K). We also derived the masses of the molecular clouds using the $W_{\rm CO}$--$N$(H$_{2}$) conversion factor ($X_{\rm CO}$ = 7.0$\times$10$^{20}$ cm$^{-2}$ (K km s$^{-1}$)$^{-1}$; \cite{2008ApJS..178...56F}). We use{d} the equation as follows,
\begin{equation}
{M{\rm (H_2)} = m_{\rm H} \mu \Omega D^2 \Sigma_i N_i({\rm H2}),}
\end{equation}
{where $m_{\rm H}$ is the mass of the hydrogen atom, $\mu$ is the mean molecular weight relative to a hydrogen atom, $D$ is the distance to the source in cm, equal to 50 kpc, $\Omega$ is the solid angle subtended by a unit grid spacing of a square pixel, and $N_i$(H$_2$) is the hydrogen molecule column density for each pixel in unit of cm$^{-2}$. We adopt $\mu$ = 2.7 to take into account the $\sim$36\% abundance by mass of helium relative to hydrogen molecule. }
\begin{equation}
{N(H_{2})=X_{\rm CO}\times W (^{12}CO(J=1-0)), }
\end{equation}
where $W_{\rm CO}$ is the integrated intensity of $^{12}$CO($J$=1--0) and $N$(H$_{2}$) is the column density of molecular hydrogen. The masses of atomic hydrogen and molecular hydrogen are 0.3$\times$10$^8$ $M_{\rm \odot}$ and 0.3$\times$10$^7$ $M_{\rm \odot}$ for the L-component, 0.8$\times$10$^7$ $M_{\rm \odot}$ and 0.9$\times$10$^7$ $M_{\rm \odot}$ for the I-component, and 1.8$\times$10$^8$ $M_{\rm \odot}$ and 2.0$\times$10$^7$ $M_{\rm \odot}$ for the D-component. {The I-component {likely} consists of the mass converted {from the L- and D-components.}

{
{We calculated {the} molecular mass fraction ($f_{\rm mol}$) of the H{\sc i} Ridge region with
\begin{equation}
    f_{\rm mol}=M({\rm H_{2}})/M({\rm HI}). 
 \end{equation}
 $f_{\rm mol}$ of the L-, I-, and D-components is 15\%, 30\%, and 9\%, respectively. $f_{\rm mol}$ of the I-component is enhanced and is three times higher than that of the D-component {in the H{\sc i} Ridge}. This suggests that the I-component of the H{\sc i} Ridge region and molecular clouds are effectively formed in the H{\sc i} Ridge region. }}

 \begin{table}
 \begin{center}
  \caption{Mass of hydrogen gas over the whole LMC}
         \label{tab:mass}
         \begin{tabular}{ccccc}
         \hline \hline
          \multirow{2}{*}{HI component}& &M(H{\sc i})& $M$(H$_{2}$) & $f_{\rm mol}$\\
          &&{[}$M_{\odot}${]}&{[}$M_{\odot}${]}&{[}$\%${]}\\ \hline
 \multirow{3}{*}{The whole LMC}& L-comp.& 0.3$\times$10$^8$& 0.3$\times$10$^7$&$\sim$10  \\
 &I-comp.& 0.8$\times$10$^8$& 0.9$\times$10$^7$&$\sim$10 \\ 
 &D-comp. & 1.8$\times$10$^8$& 2.0$\times$10$^7$&$\sim$10 \\  \hline
  \multirow{3}{*}{The H{\sc i} Ridge}&L-comp. & 2.5$\times$10$^7$& 3.8$\times$10$^6$&$\sim$15  \\
&I-comp. & 3.4$\times$10$^7$& 1.1$\times$10$^7$&$\sim$30\\  
&D-comp. & 4.6$\times$10$^7$& 4.3$\times$10$^6$&$\sim$9 \\  \hline
\end{tabular}
    \end{center}
   \end{table}

Figure \ref{fig4} shows the {first} moment {distributions} of the three components. {The first} moment is the intensity-weighted velocity following the equation of $\Sigma$($I$ $\times$ $v$)/$\Sigma$($I$), where $I$ is the intensity of emission and $v$ is $V_{\rm offset}${, which is defined as $V_{\rm offset}$ = $V_{\rm LSR}$$-$$V_{\rm LSR}$ (the D-component) (Papers I and II).} Figure \ref{fig4}(a) shows the first moment of the L-component. For the H{\sc i} Ridge region, there is a velocity gradient from the east to the west. {The} typical velocity of the eastern side is $-$60 km s$^{-1}$, {which} increases to $-$30 km s$^{-1}$ on the western side. 
Figure \ref{fig4}(b) shows the {first} moment of the D-component, {where no systematic }change is found. 
Figure \ref{fig4}(c) {indicates the first moment of the I-component. The first moment of the I-component exhibits lower velocities, particularly in regions where it spatially overlaps with the L-component.}

{Figure \ref{fig4}(d) shows histograms of the first moment of the I-component. The median value of the first moment is $\sim$$-$20.5 km s$^{-1}$ at the positions where the integrated intensity of the L-component is {more extensive} than 300 K km s$^{-1}$. Meanwhile, the median of the first moment toward the other regions is $\sim$$-$16.5 km s$^{-1}$. Thus, the blue-shifted velocity of the I-component is affected by the dense part of the L-component{, and the red-shifted velocity of the I-component is affected by the D-component}. This comparison indicates that the I-component is strongly influenced by the L-component, {which} is consistent with the fact that the I-component is induced by the interaction driven by the L-component.}

\begin{figure*}[htbp]
\begin{center}
\includegraphics[width=\linewidth]{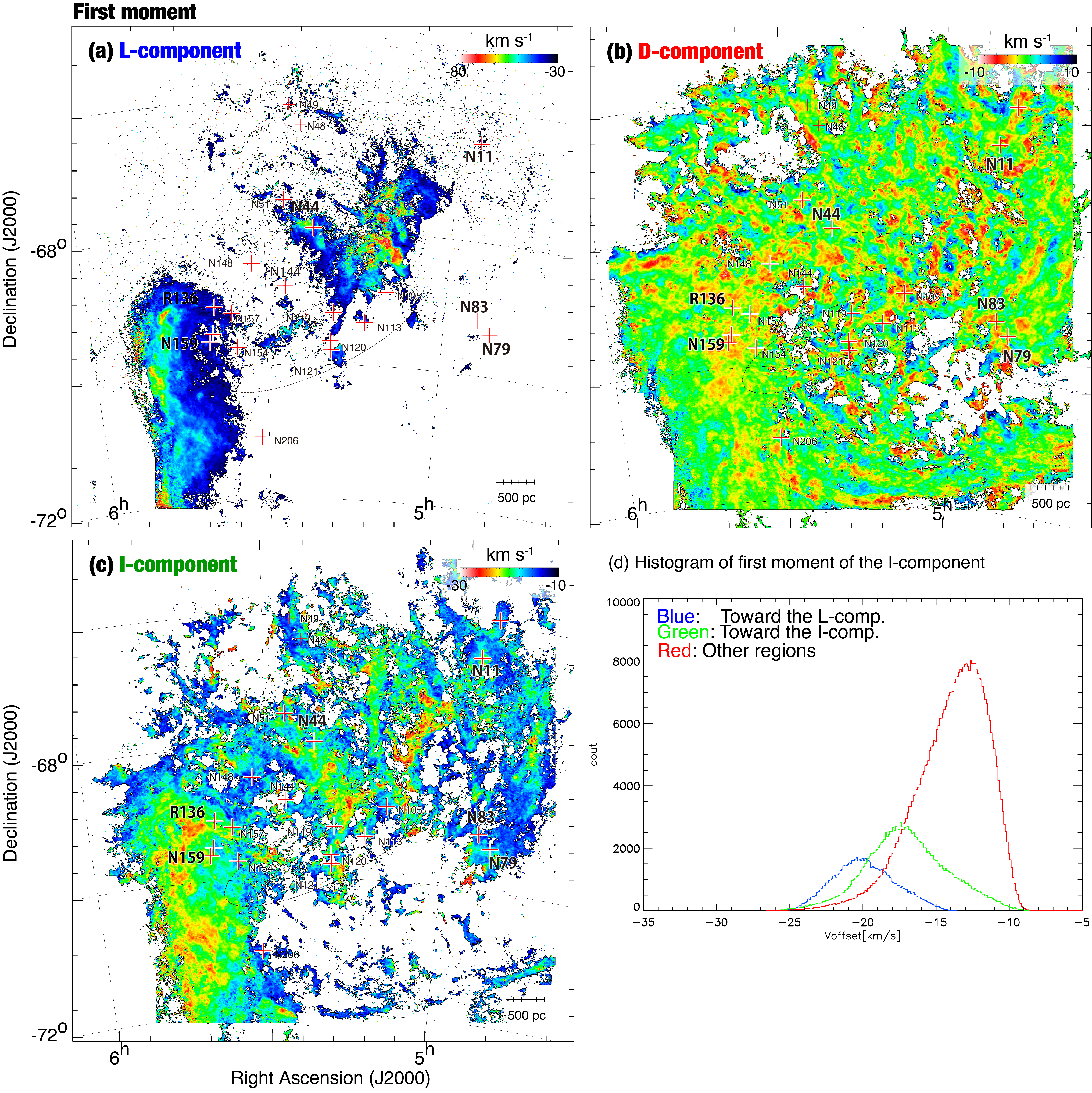}
\end{center}
\caption{First moment map of H{\sc i} over the whole LMC. Velocity ranges using for calculating momentum are $V_{\rm offset}$=$-101.1$--$-$30.5 km s$^{-1}$ (the L-component) for (a); $V_{\rm offset}$=$-10.4$--$-$9.7 km s$^{-1}$ (the D-component) for (b); $V_{\rm offset}$=$-30.5$--$-$10.4 km s$^{-1}$ (the I-component) for (c). The symbols are the same as in Figure 1. {(d) Histograms of first-moment map of the I-component. The histograms of the first moment at the positions where the integrated intensity of the L-component is larger than 300 K km s$^{-1}$ are shown in blue. The histograms of the first moment at the positions where the integrated intensity of the I-component is larger than 300 K km s$^{-1}$ are shown in green. The histograms of other regions {which is dominated by the D-component} are shown in red. {Dashed lines show the median value of the first-moment map.}}}  
\label{fig4}
\end{figure*}%

\subsection{Comparison of the I-component and high-mass stars}\label{sec:3.3}

We find by eye inspection that the I-component shows {the} best association with the major star-forming regions among the three. This motivates us to explore further details of the association with the high-mass stars {with} the I-component in the following. In Figure \ref{fig5}(a), we compare the spatial distributions of the I-component and 697 O/WR stars (hereafter O/WR stars) \citep{2009AJ....138.1003B}. These O/WR stars allow us to examine the correlation more extensively than the major star-forming regions. In Figure \ref{fig5}(b), we also show the distribution of the H$\alpha$ emission overlaid on the I-component, which traces the effect of the ionization/feedback on the H{\sc i} by the O/WR stars. 

{T}o inspect the feedback effects due to ionization{/stellar winds} by the O/WR stars, we compare spatial distributions of H$\alpha$ emission and H{\sc i} gas toward the three regions around {the }R136, N11, and {N77-N79-N83 complex} at a 10--100 pc scale. Figures \ref{fig5}(c), \ref{fig5}(d), and \ref{fig5}(e) show enlarged views of {the} H{\sc i} toward N11, R136, and {N77-N79-N83 complex}, respectively. We present detailed velocity channel maps toward R136, N11, and {N77-N79-N83 complex, which} are shown in Figures \ref{fig17}, \ref{fig18}, and \ref{fig19} of Appendix 2, and find the H{\sc i} intensity depression toward the H{\sc ii} region {often} in the velocity range of the I-component {($V_{\rm offset}$ = $-$30.5--$-$10.4 km s$^{-1}$)}. These depressions are consistent with the stellar feedback effects. In R136, the depression is {partly} due to H{\sc i} absorption of the radio continuum emission of the H{\sc ii} region. {N79 (Fig{ure} 16) is a young star-forming region (Ochsendorf et al. 2017), so the H{\sc i} depression is not prominent.}

\begin{figure*}[htbp]
\begin{center}
\includegraphics[width=\linewidth]{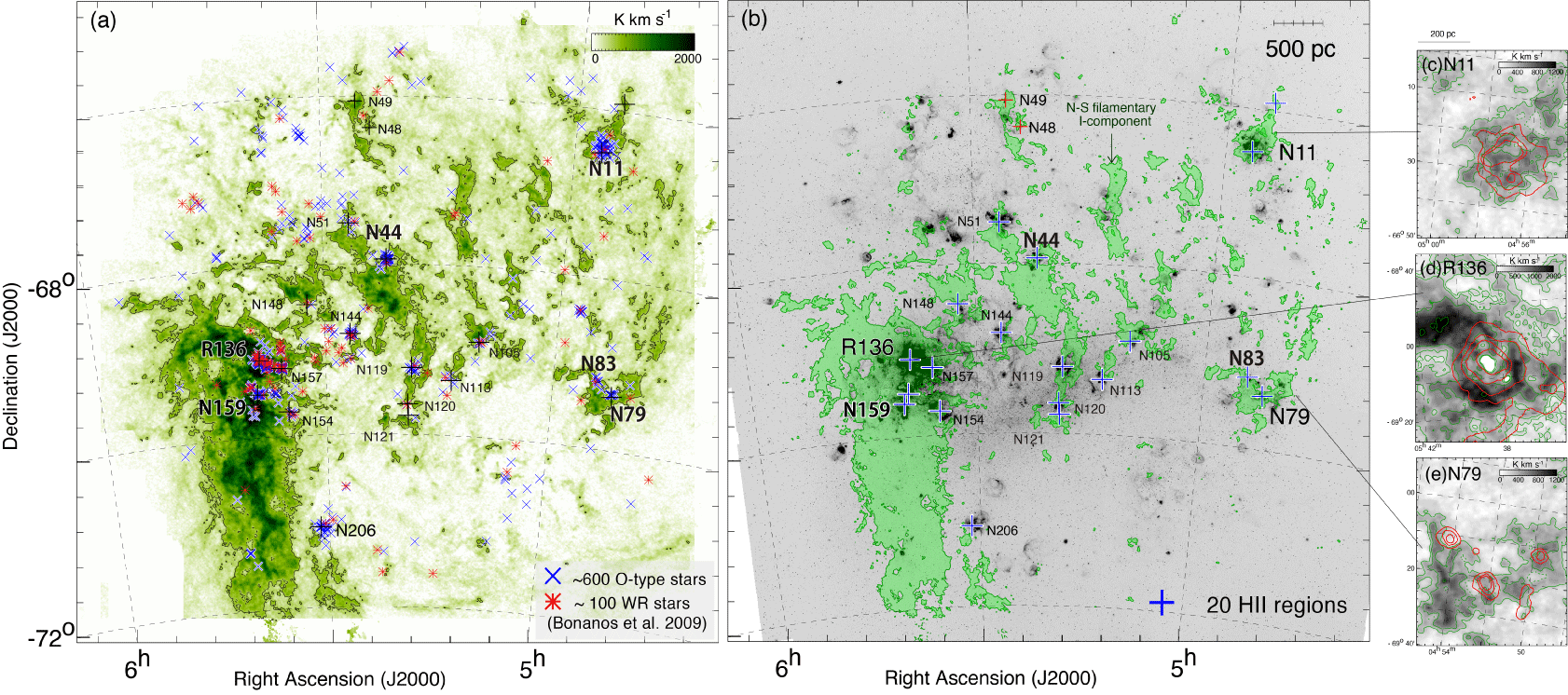}
\end{center}
\caption{(a) The H{\sc i} integrated intensity map of the I-component. The red asterisks and blue crosses indicate WR stars and O-type stars (Bonanos et al. 2009), respectively. (b) The H{\sc i} integrated intensity map of the I-component by contours superposed on H$\alpha$ image. The symbols are the same as in Figure \ref{fig1}. The contour level of H{\sc i} is 300 K km s$^{-1}$. (c) (d) (e) show enlarged view toward N11, R136, and N79, respectively. Images and green contours show H{\sc i} distributions. Red contours show H$\alpha$ image. The contour levels of H{\sc i} are 50, 150, and 250 K km s$^{-1}$. The contour levels of H$\alpha$ are 150, 300, 450, and 600 R. Detailed velocity channel maps are shown in Appendix 2. }  
\label{fig5}
\end{figure*}%

{To quantify the spatial correlation between the O/WR stars and the I-component,} we {present} a histogram of {the} integrated intensity of the I-component (hereafter, $W_{\rm HI}$ (I)) at the positions of 697 O/WR stars cataloged by \cite{2009AJ....138.1003B} as shown in Figure \ref{fig6}. {This figure indicates } that $\sim$50\% of the O/WR stars are located at positions where $W_{\rm HI}$ (I) $>$ 300 K km s$^{-1}$ (green histogram of Figure \ref{fig6}(a)). This correlation cannot be due to chance coincidence, because the characteristics of the histogram {are} significantly different from what is expected for the case of a purely random distribution, {which is} shown by the grey histogram {in} Figure \ref{fig6}(a). {The H{\sc i} intensity depressions inspected above in {the} R136, N11, and {N77-N79-N83 complex} suggest that $W_{\rm HI}$ is probably decreased by the stellar feedback toward the H{\sc ii} regions from the initial value {before} the star formation. For the depressions of $W_{\rm HI}$ distribution, which have radii around 50 pc in Figure \ref{fig5}, we corrected the value of $W_{\rm HI}$(I) for the feedback effects if more than one pixel has $W_{\rm HI}$ (I) smaller than 300 K km s$^{-1}$ i) and H$\alpha$ emission higher than 500 {deci Rayleigh (dR)} ii). 50 pc is a radius expected for an ionized cavity by an O star in 10 Myr when the velocity of the ionization front is assumed to be 5 km s$^{-1}$ (e.g., \cite{2018ApJ...859..166F}). In the correction, we replace $W_{\rm HI}$ toward the star {with} the highest value of $W_{\rm HI}$(I) within 50 pc of the star. This method is not so strict, but we find that it is helpful to fill the obvious H{\sc i} depressions. Figure \ref{fig6}(b), the histogram corrected for the H{\sc i} depression, indicates that {519/697} (74 \%) of the O/WR stars are located at positions where $W_{\rm HI}$(I) $>$ 300 K km s$^{-1}$. The blue histogram of Figure \ref{fig6}(b)) {is different from the random case (grey) and becomes more significant than in Figure \ref{fig6}(a) without correction.} }
{We shall discuss {the} possible implications of the correlation between the O/WR stars and the supergiant shells in Section 5. }
{Figures \ref{fig7}--\ref{fig10} show detailed H{\sc i} data of the H{\sc i} Ridge, N11, and {N77-N79-N83 complex} regions.
We defer {complete} discussion on these figures in Section 5 by considering the new numerical simulations presented in Section 4.}

\begin{figure}[htbp]
\begin{center}
\includegraphics[width=\linewidth]{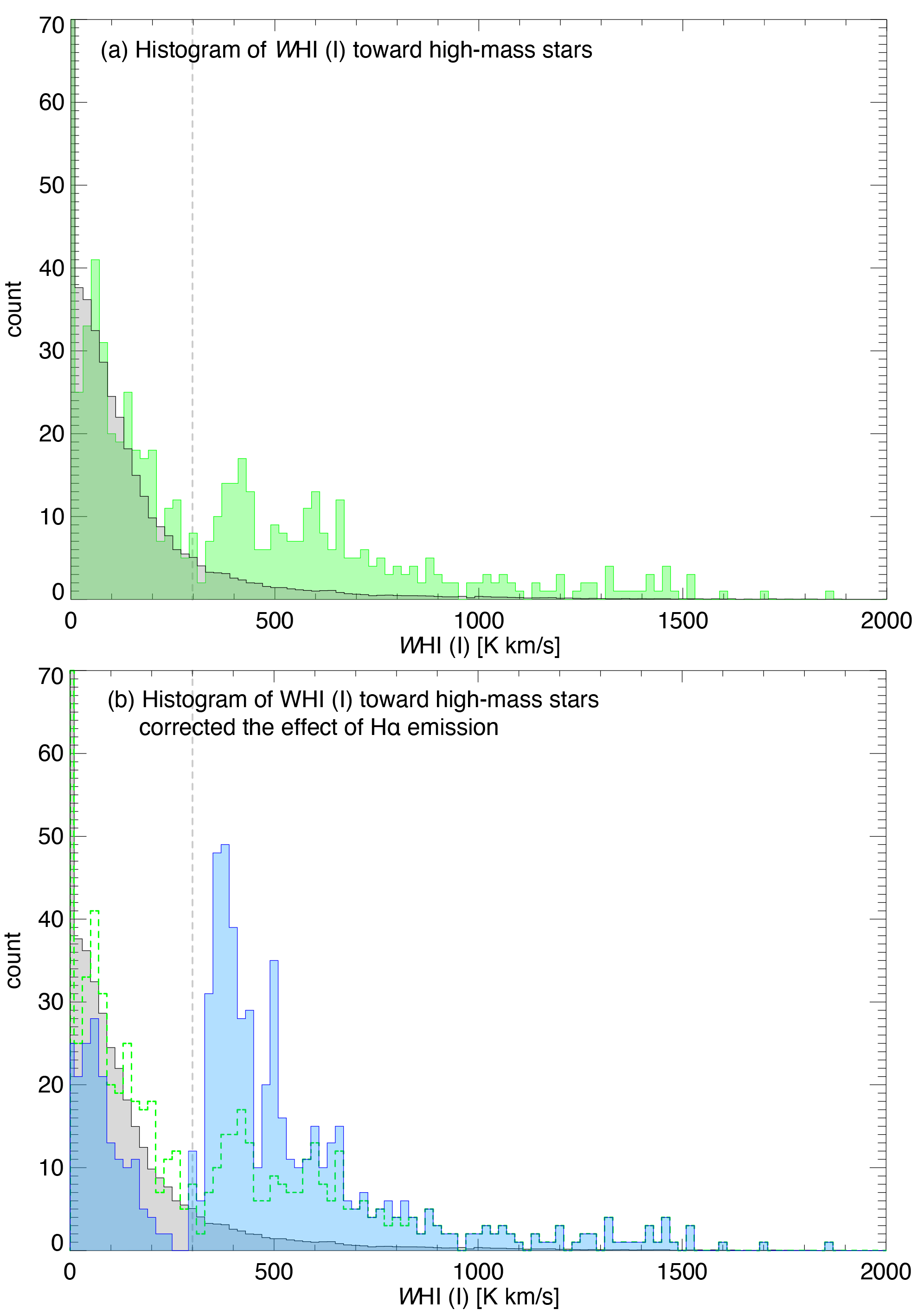}
\end{center}
\caption{(a) Histograms of integrated intensity of the I-component ($W_{\rm HI}$ (I)) at the positions of 697 O/WR stars cataloged by Bonanos et al. (2009) shown in green. Grey histogram shows {the distribution} expected if the same number of high-mass stars are distributed at random over the whole LMC. (b) Green and grey histograms are the same as in (a). {The blue histogram in (b) shows the distribution corrected for the effect of local ionization by high-mass stars. For the detailed method see section 3.{3}.{The black vertical dashed lines indicate the threshold of ionization correction, which is set at 300 K km s$^{-1}$}}}  
\label{fig6}
\end{figure}

\begin{figure*}[htbp]
\begin{center}
\includegraphics[width=\linewidth]{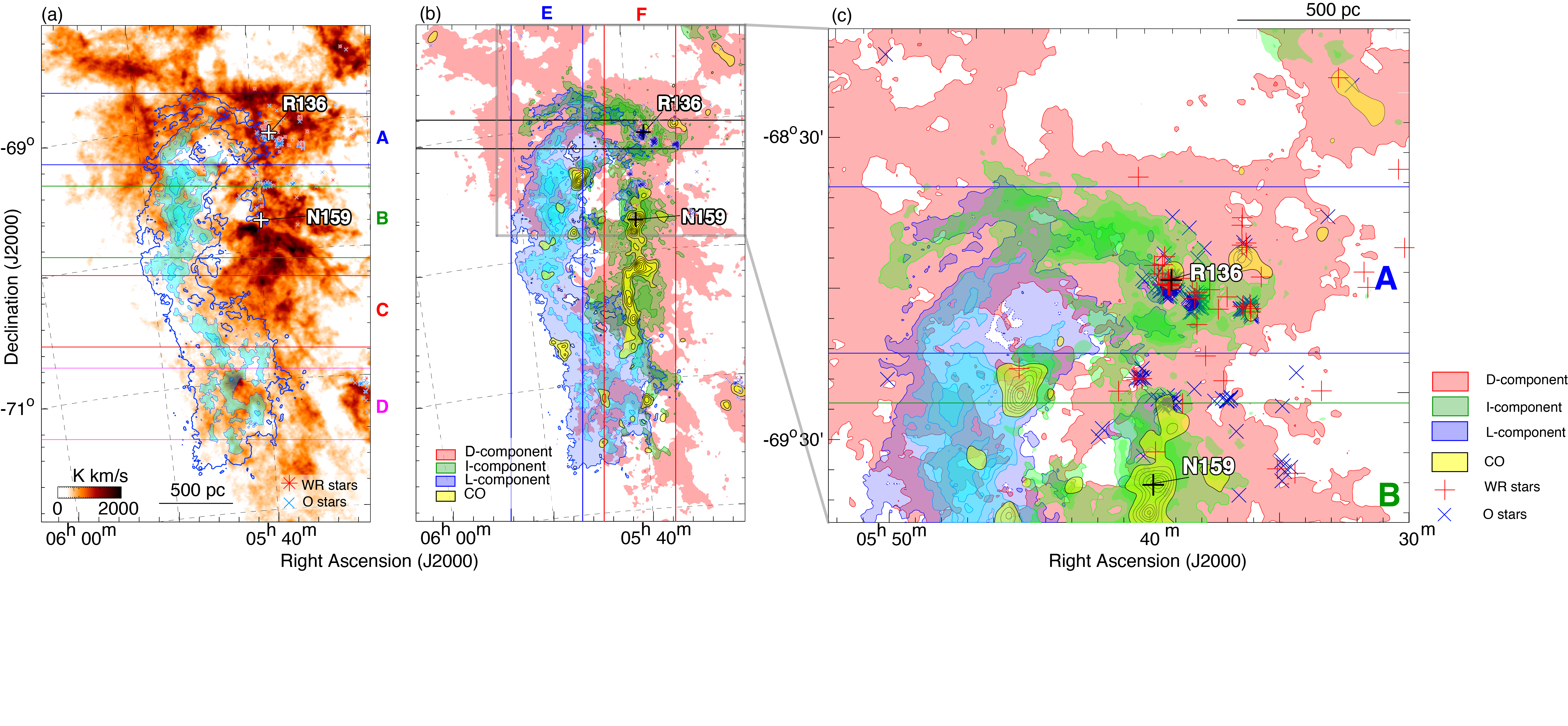}
\end{center}
\caption{{(a) Intensity map of H{\sc i} consisting of the L- and D-components {in} the H{\sc i} Ridge. The intensity map of the L-component by blue contours superposed on the D-component (orange image). The contour levels are 500, 800, 1000, 1200, 1400, 1600, 1800, and 1900 K km s$^{-1}$. Red asterisks and light blue crosses indicate WR stars and O-type stars (Bonanos et al. 2009), respectively. The line A, line B, line C, and line D show the range in Declination where Figure 7a, 7b, 7c, 7d were made, respectively. (b) Intensity map of H{\sc i} consisting of {the} three velocity components (the L-, I-, and D-components) in the H{\sc i} Ridge.The red shaded area indicates the D-component where integrated intensity is larger than 500 K km s$^{-1}$. The intensity map of the L-component by blue shaded areas and contours. The contour levels are the same as in (a). The green shaded area and contours indicate the I-component where integrated intensity is larger than 800 K km s$^{-1}$. The contour levels are 800, 1200, 1600, and 2000 K km s$^{-1}$.  The yellow shaded area and black contours indicate the distributions of $^{12}$CO($J$=1--0) obtained {with the} NANTEN telescope. The integration velocity range is $V_{\rm offset}$=$-101.1$--9.7 km s$^{-1}$. The lowest contour level and intervals are 1.48 K km s$^{-1}$ (1.5 $\sigma$). Line E and line F show the range in R.A. where Figure 7e and 7f were made, respectively. The two horizontal black lines show the range in Dec. where Figure 9a was made. (c) An enlarged view of the northern part the H{\sc i} Ridge. The red crosses and blue crosses indicate WR stars and O-type stars (Bonanos et al. 2009), respectively. }}  
\label{fig7}
\end{figure*}%

\section{Discussion}\label{sec:discussion』}

\subsection{HI collision and star formation}\label{sec:5.2}
\subsubsection{{The} HI Ridge region}\label{sec:5.3.1}
{Paper I argued that the collision between the L- and D-components in the H{\sc i} Ridge compressed the gas to form R136 and nearby high-mass stars in the compressed layer. The metallicity was derived to be about 0.2 $Z_{\odot}$ in the region and was interpreted as a result of low metallicity gas injection from the SMC (Paper I). In the following, we explore a comprehensive picture of the triggered star formation by using position-velocity diagrams cover{ing} most of the H{\sc i} Ridge, allowing us to derive more details than in Paper I.} {To examine the detailed velocity structure and spatial distribution, we divided the H{\sc i} Ridge region into Lines A to F and conducted analyses. Lines A to F are all set to the same width for comparability. A to D cover the H{\sc i} Ridge region from north to south at equal intervals. Lines E and F are created to cover regions with strong L-component and strong I-component, respectively.}

\paragraph{Complementary distribution} \ \\

Figure \ref{fig7}(a) shows spatial distributions of the L- and D-components toward the H{\sc i} Ridge region. We recognize that the D-component shows intensity depression toward the dense part of the L-component in Zones A and B in Figure\ref{fig7}(a) at Dec. $\sim$$-$70d--$-$69.5 d, which corresponds to the complementary distribution between the L-component and the D-component (Paper I). The complementary distribution in the north show{ed} a displacement of 260 pc in a position angle of 45 degrees and was interpreted as due to {the} motion of the L-component relative to the D-component from the northwest to the southeast (Paper I). Such a displacement is a common signature in colliding clouds \citep{2018ApJ...859..166F}. Toward Zones C and D in the south, the L- and D-components significantly overlap, and the complementary distribution is less clear. {In contrast,} some parts of the distribution may be interpreted to be complementary (Paper I).

{Figure \ref{fig7}(b) overlays the I-component on the L- and D-components. Figure \ref{fig7}c is an enlarged view of the northern H{\sc i} Ridge. The I-component is distributed mainly in the west and north of the L-component and is overlapped with the D-component. The L-component is associated with the CO-Arc in Zone E. {T}he I-component is associated with the CO Molecular Ridge in Zone F.  We recognize that high-mass star formation is active in the north of the Molecular Ridge at Dec.$>$$-$70d (Zones A--B), where R136, N159 and the other H{\sc ii} regions as well as the giant molecular clouds are distributed. The distribution of the O/WR stars shows good correspondence with the I-component particularly in the west of R136 (Figure \ref{fig7}(c)). On the other hand, in the CO Arc and the south of the Molecular Ridge at Dec.$<$$-$70d (Zones C--D), we find no active high-mass star formation in the giant molecular clouds.}

\paragraph{E-W and N-S distributions of the HI components; details of merging} \ \\

{In colliding flows, we expect {the} merging of the two velocity components, another feature characteristic {of the} collision. We explore details of the collisional merging in the H{\sc i} Ridge. }
{Figures \ref{fig8new} a--d show R.A.--velocity diagrams of H{\sc i} and CO in Zones A--D (Figure \ref{fig7}a). 

{From Figures \ref{fig8new}a–d, the P–V diagram shows a velocity gradient ranging from $-$50 km {s$^{-1}$} to 0 km {s$^{-1}$} over 500 pc from east to west.}
{Thus,} in Zones A--C, {H{\sc i} gas} shows a velocity gradient of $\sim$50 km s$^{-1}$/500 pc = 0.1 km s$^{-1}$ pc$^{-1}$ in R.A. and seems to merge with the D-component in the west at R.A.= 5$^h$ 35$^m$--45$^m$ along with the broad bridge features between the L- and D-components {as shown in Figures \ref{fig8new}a--c}. {We find a trend that the merging between the I- and L-components is more developed in the west of the L-component along the Molecular Ridge than in the east along the CO Arc. The velocity of the D-component is flattened at $V_{\rm offset}$ = 0, subtracting the rotation of the galaxy (Paper II). On the other hand, the velocity of the L-component varies with the right ascension position. Regardless of which figure (Figures \ref{fig8new}a--d) is consulted, around 5$^{\rm h}$50$^{\rm m}$, the velocity is consistently in the range of $-$50 to $-$60 km s$^{-1}$.} The velocity gradient of the L-component is a natural outcome of the collisional deceleration. {It} suggests that the collisional interaction is taking place over a large extent of the L-component (Paper I). The I-component is a result of the deceleration and is distributed between the L- and D-components at R.A. =5$^h$ 35$^m$--45$^m$ in Figures \ref{fig8new}a--d. 

{Figure \ref{fig8new}e shows a Dec.--velocity diagram in Zone E which includes most of the CO Arc. This clearly shows the L- and D-components as well as the several bridge features between them at Dec.$<$$-$69 d. In the region, the I-component is {in}significant except for the northern end at Dec.$\sim$$-$69d., where the I component becomes enhanced, and the L-component is weak. This indicates significant merging of the two components into the I-components only at Dec.$\sim$$-$69d. It is also {evident} that the CO clouds in the CO Arc are all associated with the L-component.}

{Figure \ref{fig8new}f shows a Dec.--velocity diagram in Zone F. We find that the I-component {prevails} at Dec.$>$ $-$70.5d, where the Molecular Ridge is distributed. Toward the I-component, the L-component is very weak. At Dec.$<$$-$70.5d, the I-component is weaker than at Dec.$>$ $-$70.5d and seems connected with the L- and D-components forming a few bridge features at Dec.$\sim$$-$70.5d--$-$71.5d. At Dec. of N159, the maximum H{\sc i} column density is 1.7$\times$10$^{21}$ cm$^{-2}$ in Zone E, while that in Zone F is more than doubled to 4$\times$10$^{21}$ cm$^{-2}$. This increase is consistent with the fact that the L- and D-components are merging toward the Molecular Ridge. We also note {a} moderate decrease {in} the H{\sc i} column density of the I-component from the north to the south; the H{\sc i} column density is $\sim$4$\times$10$^{21}$ cm$^{-2}$ in Zones A--B, $\sim$3$\times$10$^{21}$ cm$^{-2}$ in Zone C, and $\sim$2.5$\times$10$^{21}$ cm$^{-2}$ in Zone D, suggesting a N--S gradient in column density.}

{Figure \ref{fig8} focuses on the horizontal Zone toward the R136 region, indicated by two black lines in Figure \ref{fig7}b. 
The number of O/WR stars is plotted in Figure \ref{fig8}b, which peaked toward R136 and is enhanced in an R.A. range from 5$^h$35$^m$ to 5$^h$40$^m$. These O/WR stars are {possible} outcomes of the triggered star formation by the same event {that} formed R136 as previously suggested (Paper I). In Figure \ref{fig8}c, the four H{\sc i} profiles along the Zone show details of merging, which forms the I-component from the east to the west. Further details of the triggering process will be clarified by investigating the stellar properties, which {are} beyond the scope of the present paper.}

\paragraph{Physics of the merging of the HI components} \  \\

{{T}he above {shows} that the merging process seems different between the CO Arc and the Molecular Ridge. In the Molecular Ridge, the collision of the L- and D-components leads to merging to form the I-component and the Molecular Ridge, whereas it does not lead to merging in the CO Arc as shown by no intense I-component toward the CO Arc. We explore how the difference is explained below.}

{A possible scenario i) is that the collision toward the CO Arc is in the early stage, and the merging of the two components has not yet occurred significantly, whereas the two components collided heavily toward the present Molecular Ridge and developed the I-component. We suggest that the Molecular Ridge was formed by the H{\sc i} collision in the I-component in less than 10 Myr, while the molecular gas in the CO Arc was pre-existent {before} the collision. This is possible if the initial separation between the L-component and the D-component is {more prominent} toward the CO Arc than toward the Molecular Ridge. We suggest a tilt of the L-component in the east to west relative to the D-component explains such a scenario if we assume that the L-component {is} a flat plane-like cloud. An alternative scenario ii) is that the L-component {of} the CO Arc region has significantly higher H{\sc i} column density than that of the D-component. {In contrast,} the column density was similar between the L- and D-components toward the Molecular Ridge. {In scenario ii)}, the L-component {in} the CO Arc {experiences minimal deceleration even upon collision.} Because the CO Arc is located toward the edge of the LMC disk, the lower column density of the D-component is reasonable. At Dec.$\sim$$-$70.3d.--$-$69.3d the column density of the L-component, which is associated with the giant molecular clouds, is higher than that of the D-component (Figure \ref{fig8new}e) and may support scenario ii). In summary, we find that the two scenarios are viable explanations. They are not exclusive and both may be working.}

\paragraph{Overall high-mass star formation in the HI Ridge} \  

{The formation of the high-mass stars and the GMCs is active only in the northern half of the H{\sc i} Ridge (Figure \ref{fig7}b); $\sim$400 O/WR stars {are} concentrated in an area of $\sim$500 pc $\times$ $\sim$500 pc in the northwest of the H{\sc i} Ridge at Dec.$>$$-$70.8 d, and the GMCs are distributed at Dec.$>$ $-$70.0 d. On the other hand, the southern half of the H{\sc i} Ridge is quiescent in high-mass star formation. {The morphology that the absorption in the soft X-rays independently supports the L-component with a tilt located in front of the LMC disk (Sasaki et al. 2022; Knies et al. 2021) and in the near infrared extinction (Furuta et al. 2019, 2021). It is worth mentioning that the soft X-rays are likely emitted from the gas between the Molecular Ridge and the CO Arc heated by the gas collision at $\sim$100 km s$^{-1}$ (Knies et al. 2021) as is consistent with the colliding H{\sc i} flows. This direction of the falling motion of the L-component is further supported by the ALMA observations of colliding CO clouds in N159 (Fukui et al. 2019; Tokuda et al. 2019; 2022).}
\begin{figure*}[htbp]
\begin{center}
\includegraphics[width=\linewidth]{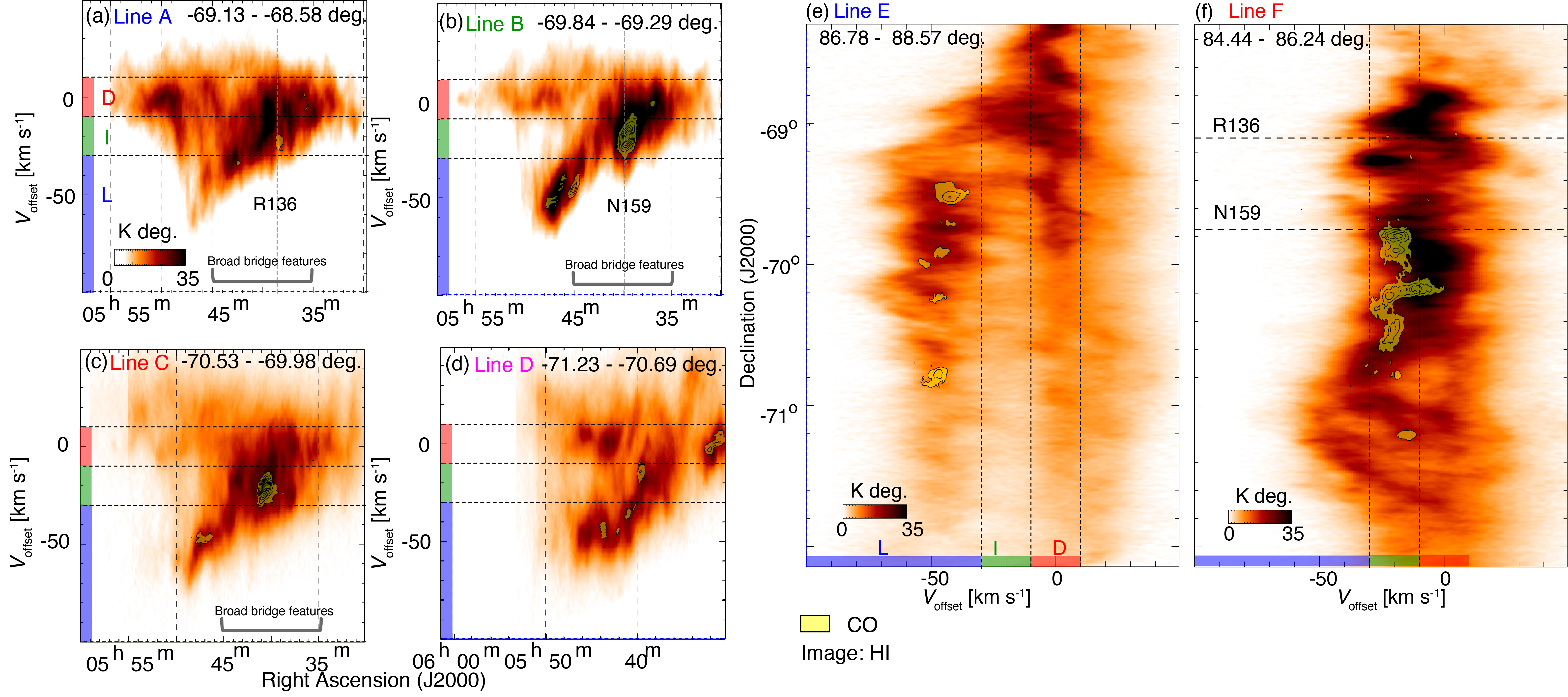}
\end{center}
\caption{{(a)--(b) Right Ascension--velocity diagrams of H{\sc i} by image overlaid with the CO by yellow contours along lines A, B, C, and D in unit of K degree. The integration is in Dec. over the range from $-$69.13 to $-$68.58 deg. for (a), $-$69.84 to $-$69.29 deg. for (b), $-$70.50 to $-$69.95 deg. for (c), and $-$71.32 to $-$70.77 deg. for (d). The lowest contour level and intervals are 0.0241 K deg. and 0.0241 K deg., respectively. (e)(f) Declination--velocity diagram of H{\sc i} by image overlaid with the CO by yellow contours along Lines E and F in unit of K degree. The integration ranges in Right Ascension are from 86.78 to 88.57 degree for (e) and 84.44 to 86.24 degree for (f). The lowest contour level and intervals are 0.0241 K deg. and 0.0241 K deg., respectively.}}  
\label{fig8new}
\end{figure*}%
{These observational trends obtained with H{\sc i}, CO, and X-ray} suggest that the collisional compression propagates from the north to the south, and the active high-mass star formation only in the north may be explained by the propagating time-dependent star formation in a timescale of $\sim$10 Myr.} {Figure \ref{figgeo} shows {the} 3D geometry of the collision toward the H{\sc i} Ridge}. The time scale is roughly consistent with the evolutionary stage of giant molecular clouds (GMCs) of the H{\sc i} Ridge region (\cite{1999PASJ...51..745F,2009ApJS..184....1K}, see also Kawamura 2010). In addition, it is interesting to note that the GMC evolutionary stages show a north-south sequence of star formation. In the northern part, most of the molecular gas around R136 {is} classified as Type III, a molecular cloud associated with active cluster formation and H{\sc ii} regions. In the southern part of R136 including N159, there are many Type II GMCs, which are in the younger stage {and are} associated only with H{\sc ii} regions. In the more southern region, the youngest Type I GMCs without associated H{\sc ii} regions are dominant. The time scales of Type I, Type II, and Type III are estimated to be 6 Myr, 13 Myr, and 7 Myr, respectively, by \cite{2009ApJS..184....1K}, which are similar to the time scale of the collision. Thus, the north-south sequence of star formation in the H{\sc i} Ridge region may be explained by the three-dimensional structure of the collision we proposed.

{We need to consider the smaller-scale processes to deepen our understanding of the high-mass star formation in the H{\sc i} Ridge. In particular, the recent ALMA results revealed that the high-mass star formations in the ''peacock-shaped clouds'' in N159E and N159W-S are triggered by a pc-scale falling cloud colliding with extended gas (\cite{2019ApJ...886...14F,2019ApJ...886...15T}). In {addition}, such a collision {was} numerically simulated by \cite{2018PASJ...70S..53I} and it was shown that the collisional compression reproduces filamentary conical dense gas distribution. Follow-up ALMA observations revealed that N159W-N also shows the CO distribution consistent with the simulations of Inoue et al. (2018). It is notable that the direction of these falling clouds is parallel to the direction derived from the kpc-scale displacement in the NW--SE direction.
  \citep{2023ApJ...955...52T}. These results suggest that the interaction of falling clouds with the disk, as a consequence of the tidal interaction, is a vital process to form high-mass stars.} {We {present} a detailed picture that the colliding H{\sc i} flows consist of small dense (10--100 cm$^{-3}$) H{\sc i} clumps of pc-scale and the clumps trigger high-mass star formation at individual spots separated by tens of pc over a kpc scale. Future ALMA observations will be crucial to broaden the application of the scenario to the rest of the LMC where high-mass star formation is occurring. }

\subsubsection{The Diffuse L-component}\label{sec:5.4.2}
In Paper II, the colliding H{\sc i} flows with low metallicity are found to trigger the high-mass star formation in N44. N44 is part of the Diffuse L-component, and it is likely that the colliding H{\sc i} flows include the metal-poor gas injected from the SMC, as shown by the dust-to-gas ratio in Paper II. 
\begin{figure*}[htbp]
\begin{center}
\includegraphics[width=12cm]{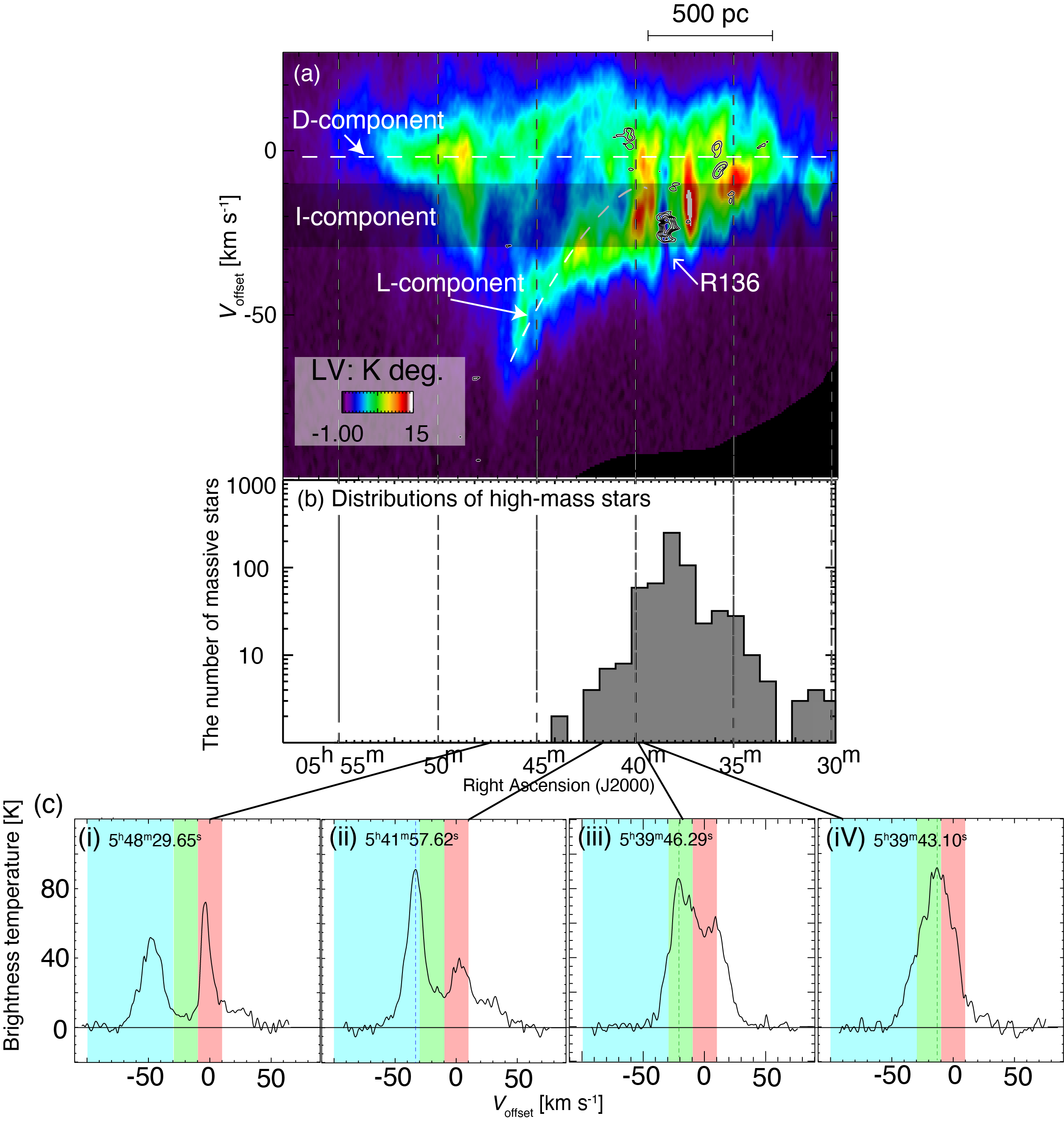}
\end{center}
\caption{(a) Right Ascension--velocity diagram toward the northern part of H{\sc i} Ridge. Integration range in Declination is from $-$69.01 deg. to $-$68.79deg. at R.A. of 6h 00m 00s as shown {by two} black lines in Figure 7(b). The black contours indicate $^{12}$CO ($J$=1--0). The contour levels are 0.0133, 0.0171, 0.0209, 0.0247, and 0.0285 K deg. (b) Histogram of the {number of the} high-mass stars of {the} H{\sc i} Ridge in the same field as in Figure 7({c}). The horizontal and vertical axes are R.A. and the number of massive stars located within each R.A. bin, respectively. (c) Typical spectra of H{\sc i} at the position of (i) (R.A., Dec.)=(5$^{h}$48$^{m}$29.65$^{s}$,-69$^{d}$23'3.49"), (ii) (R.A., Dec.)=(5$^{h}$41$^{m}$57.62$^{s}$,-69$^{d}$09'26.54"), (iii) (R.A., Dec.)=(5$^{h}$39$^{m}$46.29$^{s}$,-69$^{d}$09'42.95"), and (iV) (R.A., Dec.)=(5$^{h}$39$^{m}$43.10$^{s}$,-69$^{d}$00'20.42"). Blue, green, and red shaded ranges are $V_{\rm offset}$=$-101.1$--$-$30.5 km s$^{-1}$ (the L-component), $V_{\rm offset}$=$-30.5$--$-$10.4 km s$^{-1}$ (the I-component), and $V_{\rm offset}$=$-10.4$--$-$9.7 km s$^{-1}$ (the D-component), respectively.}  
\label{fig8}
\end{figure*}%
The H{\sc i} position--velocity diagrams in Figures \ref{fig21} (a) to (k) of Appendix 4 show that the L- and D-components are connected in a velocity space toward N44. 

{The Diffuse L-component has an approximate size of 1.5 kpc $\times$1.5 kpc in R.A. and Dec. with a triangle shape whose vertex is directed toward the south (Figure \ref{fig3}). The I-component toward the Diffuse L-component is divided into four features; they are the southeastern part, including N44, N51, and N144, the southwestern part, including N105 and N113, the southern part, including N119, N120, and N121, and the northern filamentary part with a few small H{\sc ii} regions and no major H{\sc ii} regions (Figure \ref{fig5}b).

\begin{figure*}[htbp]
\begin{center}
\includegraphics[width=\linewidth]{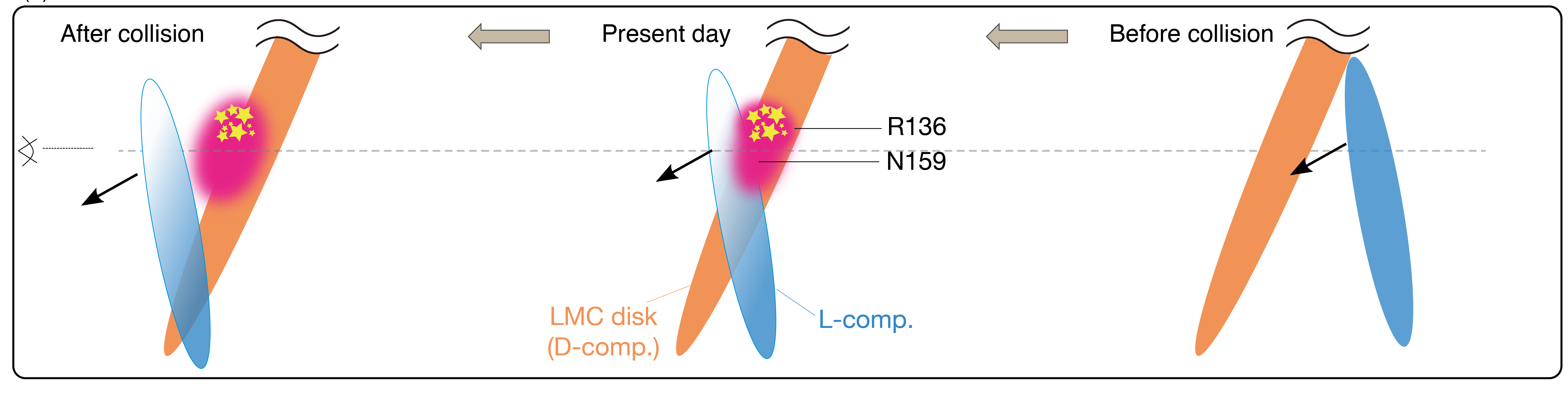}
\end{center}
\caption{{Schematic view of the collision. Orange and blue disks are the D- and L-component, respectively. The red area represents the distribution of diffuse soft X-ray (e.g., Knies, et al. 2021). The line of sight direction is from left to right. From right to left, it shows the geometry before the collision $(right)$, the present state $(center)$, and the geometry after the collision $(left)$.}}  
\label{figgeo}
\end{figure*}%

We interpret that the I-component was formed by the southward motion mainly at the southern edge of the L-component by the collisional compression, which explains that most of the I-component is distributed on the south side of the L-component. Moreover, the first-moment map shows that the Diffuse L-component is decelerated at the southern edge, as {illustrated} in Figure \ref{fig4}(a). It is possible that the Diffuse L- and D-components are merging to form the I-component in the south of the Diffuse L-component. On the other hand, the northern part of the Diffuse L-component is not decelerated and the first moment is $\sim$$-$50 to $-$60 km s$^{-1}$. There is no I-component toward this region, as shown in Figure \ref{fig3}, so the L-component is possibly {before} or at the beginning of collision {with slight deceleration}. Moreover, there is no significant molecular cloud/high-mass star formation, which is consistent with the idea that no significant compression by collision is yet taking place.}

{We suggest that in N44, the V-shaped southern part of the I-component is the strongly compressed layer that formed seven of the 20 major H{\sc ii} regions. The morphology may be explained by a scale-up version of the colliding cloud of pc-scale modeled and simulated by \citet{2018PASJ...70S..53I}, which assumes that a test spherical cloud is collid{es} with extended gas. The simulations by \citet{2018PASJ...70S..53I} show that the collision forms a compressed layer of a conical shape pointing toward the moving direction of the spherical cloud. Such conical clouds are indeed discovered in three regions of collision-induced high-mass star formation at a pc-scale in N159, i.e., N159E, N159W-S, and N159W-N (Fukui et al. 2019; Tokuda et al. 2019; 2022). These three cases are believed to be those for which Inoue et al. (2018)’s model is applicable. In the Diffuse L-component, we suggest that a kpc-scale cloud of the L-component largely collided from the north with the D-component, and the I-component of a V-shape was formed in a kpc scale. 
This scenario will be tested in more detail by using the new simulations and the H{\sc i}/CO observations toward the individual H{\sc ii} regions in the future.}

\begin{figure*}[htbp]
\begin{center}
\includegraphics[width=14cm]{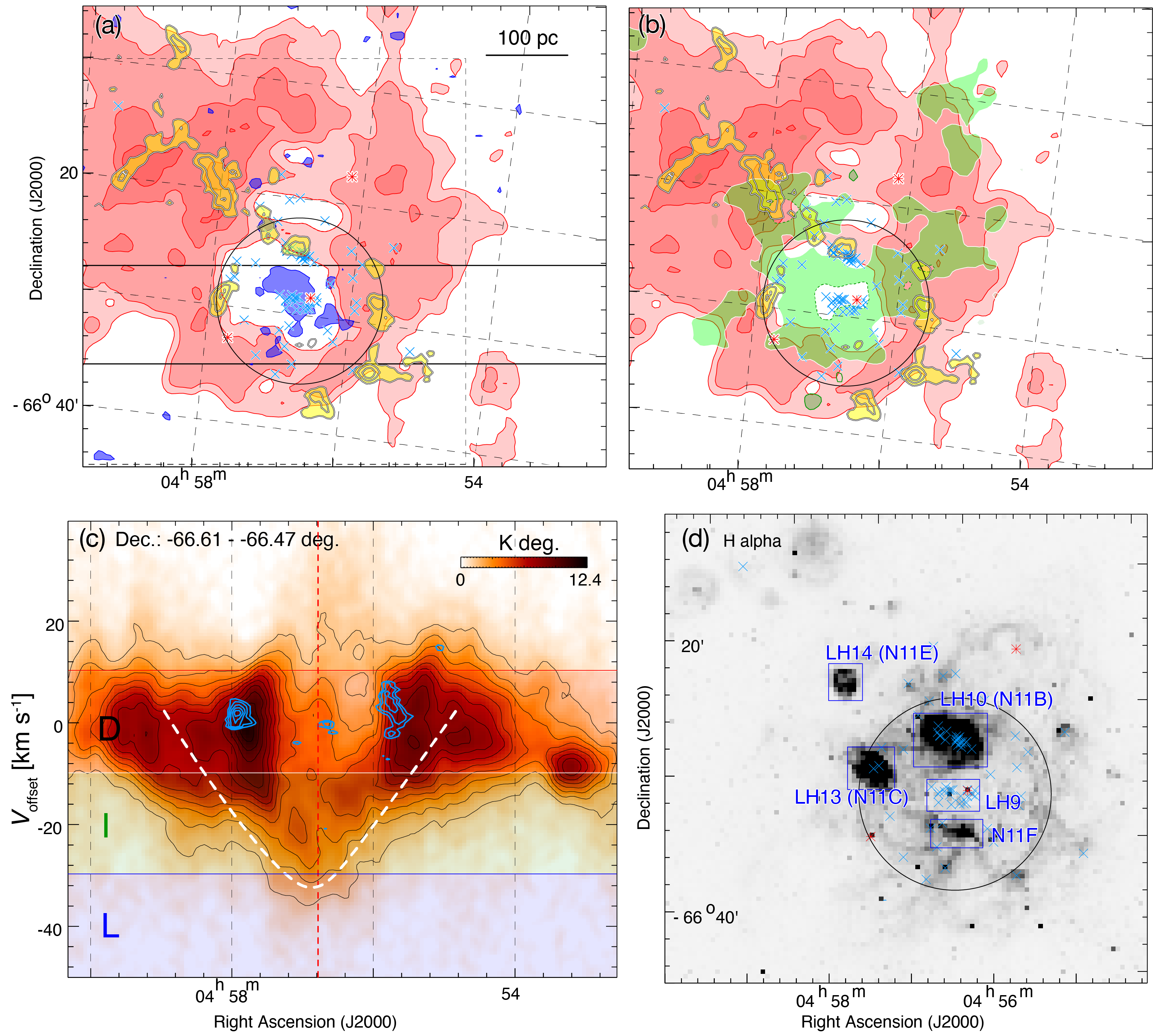}
\end{center}
\caption{(a) H{\sc i} integrated intensity map of the L-component by blue shaded area ($>$ 150 K km s$^{-1}$) superposed on the D-component by image. (b) H{\sc i} integrated intensity map of the I-component by green shaded area ($>$ 500 K km s$^{-1}$) superposed on the D-component. The white shaded area and contours of (a) and (b) indicate the total integrated intensity map of CO ($>$ 1.5 $\sigma$ (3.89 K km s$^{-1}$); Wong et al. 2011). The contour levels are 3.89, 7.78, and 12.97 K km s$^{-1}$. The red cross, red asterisks, and blue crosses indicate the position of N11 (Henize 1956), WR stars, and O-type stars (Bonanos et al. 2009), respectively. (c) Position--velocity diagram of H{\sc i} by image overlaid with the CO by blue contours. The lowest contour level and interval are 1.5 K deg. and 1.0 K deg. for H{\sc i}; 3 $\sigma$ (0.024 K deg.) and 1 $\sigma$ (0.008 K deg.) for CO.The integration range in Dec. is from $-$66.61$^{\circ}$ to $-$66.47$^{\circ}$ shown by the black horizontal lines of (a). The red dashed perpendicular line indicates the position of N11 in the R.A.. (d) H$\alpha$ image obtained by the Magellanic Cloud Emission-Line Survey (MCELS; Smith \& MCELS Team 1999) toward N11 shown by black dashed box in (a). The black circle indicates a ring morphology with a cavity of $\sim$100 pc in radius, enclosing OB association LH9 (Lucke \& Hodge 1970).}
\label{fig9}
\end{figure*}%

{The maximum column density of the L-component is $\sim$2.6$\times$10$^{21}$ cm$^{-2}$, while the D-component has a column density of 3.2$\times$10$^{20}$ cm$^{-2}$, an order of magnitude smaller than that of the Diffuse L-component at the same position. It is possible that the low column density of the D-component is a result {of} the collisional acceleration which shifted the D-component to the L-component.}

{{Examining} the effects of the stellar feedback in exploring high-mass star formation {is important}. Because the energy released by the high-mass stars is large and all the region discussed above includes 40--400 O/WR stars. By adopting typical physical parameters of the stellar feedback, we calculated the cloud mass and the kinetic energy of the L-component for velocity relative to the D-component at two assumed angles of the motion 0$^{\circ}$and 45$^{\circ}$ to the sightline. Table \ref{tab:mom} lists these physical parameters in {the} N11 and {N77-N79-N83 complex}. We find that the momentum released by the stellar feedback is lower by two orders of magnitude than that required to accelerate the motion of the I-component. Furthermore, we do not find any strong enhancement of the H{\sc i} gas motion toward the O/WR stars in each of the present regions, whereas the H{\sc i} velocity components are extended spatially. This indicates gas acceleration is extended over more than a few 100 pc but is not localized toward the O/WR stars, which is not consistent with stellar feedback. The gas motion is, therefore, not likely driven by the feedback but by the large kpc-scale tidal interaction. The similar arguments on R136 and N44 in Papers I and II are consistent with the present conclusion. {In the following section,} we will not discuss stellar feedback as a cause of gas motion in the following, while ionization localized to the individual high-mass star formation is obviously important.}

\begin{figure*}[htbp]
\begin{center}
\includegraphics[width=\linewidth]{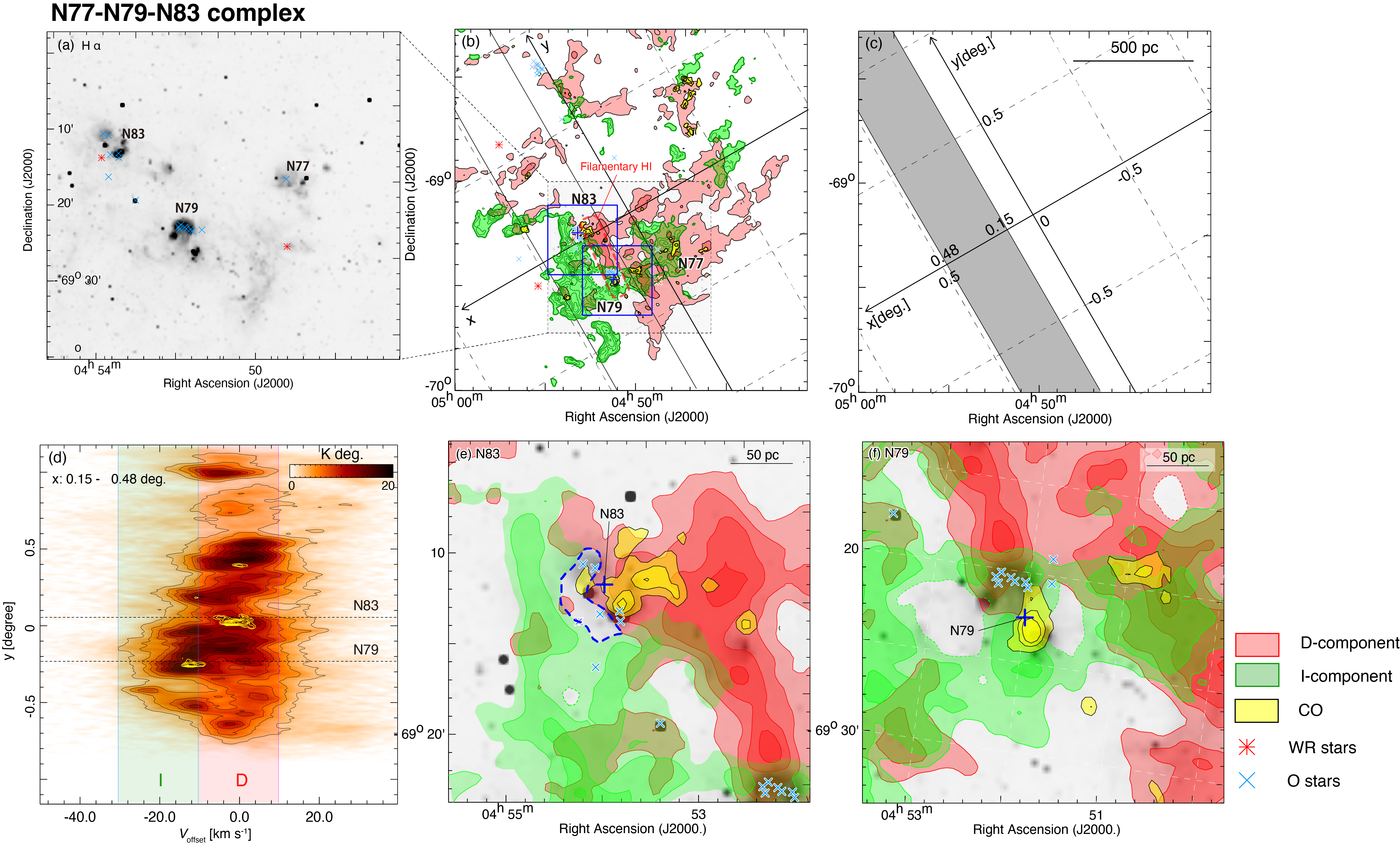}
\end{center}
\caption{ {(a) H$\alpha$ image obtained by the Magellanic Cloud Emission-Line Survey (MCELS; Smith \& MCELS Team 1999) toward {N77-N79-N83 complex} shown by black dashed box in (b). (b) H{\sc i} integrated intensity map of the I-component by green contours superposed on the D-component in red. The lowest contour level and intervals are 350 K km s$^{-1}$ and 100 K km s$^{-1}$. The yellow shaded and contours indicate the total integrated intensity map of CO ($>$ 1.5 $\sigma$ (3.89 K km s$^{-1}$); Wong et al. 2011). The contour levels are 3.89, 7.78, and 12.97 K km s$^{-1}$. The red cross, red asterisks, and blue crosses of (a), (b), (e), and (f) indicate the position of N11 (Henize 1956), WR stars, and O-type stars (Bonanos et al. 2009), respectively. (c) XY coordinate used for making position--velocity diagram shown (d). The center of the map at (R.A., Dec.)=(4$^h$51$^m$33.4$^s$, $-$69$^d$01$^m$49.3$^s$) is set as the origin of XY coordinate, and the position angle of the Y axis is $-$30 deg. (d) Position--velocity diagram of H{\sc i} overlaid with the CO contours in {yellow} toward N79. The integration range in X is from 0.15 deg. to 0.48 deg. The lowest contour level and intervals are 2.5 K deg. and 2.0 K deg. for H{\sc i} and 0.0368 K deg. and 0.0123 K deg. for CO. The positions of {N83} and {N79} are indicated by the dashed horizontal lines. (e) (f) are close-up images of (b) toward {N83} and N79, respectively. The regions are shown by blue boxes in (b)} }  
\label{fig10}
\end{figure*}%




\subsubsection{N11 region}\label{sec:4.1.3}
The H{\sc ii} region N11 was cataloged by \citet{1956ApJS....2..315H}. N11 has a ring morphology with a cavity of $\sim$100 pc in radius, enclosing OB association LH9 {in the center of the cavity } (Lucke \& Hodge 1970). {There are} several bright nebulae (N11B, N11C, N11F) around LH9 {as shown in the H$\alpha$ image (MCELS; Smith \& MCELS Team 1999) in Figure \ref{fig9}}. The massive compact cluster HD 32228 dominates this OB association and has an age of $\sim$3.5 Myr \citep{1999AJ....118.1684W}. LH 10 is the brightest nebula and lies north of LH 9. LH 10 is the youngest OB association with an age of about 1 Myr. Previous studies proposed that star formation {in} LH10 was possibly triggered by an expanding supershell blown by LH9 (e.g., \cite{1996A&A...308..588R, 2003AJ....125.1940B, 2006AJ....132.2653H, 2019A&A...628A..96C}). \citet{1996A&A...308..588R} conducted a detailed analysis of the kinematics of H$\alpha$ emission and determined an expansion velocity of 45 km s$^{-1}$ for the central cavity. The authors suggested that a dynamical age of expansion is 2.5$\times$10$^6$ years, and a possible product of the explosion includes three SNe and shock-induced star formation.

{The present H{\sc i} data reveal features {that} were not considered previously.} {Figure \ref{fig9}{a} shows the L-component and the D-component, and Figure \ref{fig9}c shows the I-component and the D-component. Figure \ref{fig9}d presents a position-velocity diagram in the Zone shown by two black straight lines in Figure \ref{fig9}{a}. The I-component and the D-component show complementary distribution. {The I-component fits} the cavity of the D-component {toward the center of the H$\alpha$ cavity}. In addition, there are O stars, emission nebulae, and GMCs outside the cavity of N11, some of which overlap with the I-component. Further, the position-velocity diagram shows that the I-component and D-component are merged, forming a V-shaped distribution, another signature of a cloud-cloud collision along with the complementary distribution \citep{2021PASJ...73S...1F}. The L-component here is a minor feature and we interpret it as part of the I-component.}

{Based on these data, we propose a colliding flow scenario as a trigger of N11, where a single spherical H{\sc i} cloud of $\sim$100pc radius collided with the D-component. The spherical cloud accompanies a few extended components of 100--200 pc length outside a 100 pc radius, which is observed as the present I-component. We assume that the collisional interaction has a path length of $\sim$100 pc, the same {as} the approximate radius of the high-mass star distribution in Figure \ref{fig9}. Then, the collision time scale is roughly estimated to be 100 pc / 30 km/s $\sim$ 3 Myr, which is consistent {with an age of LH9 (~3.5 Myr; Walborn et al. 1999)}. In the collision scenario, the age sequence between LH9 and the rest of the OB associated is ascribed to the different epochs of the collision due to the H{\sc i} morphology. First, a collision triggered star formation in LH9 at the center of N11. {T}hen, the collision proceeded outward in a shell-like distribution and triggered the formation of the other younger OB associations in the periphery, including LH11, LH10, and LH13 in the last few Myr. Additionally, LH13 was formed by the collision between the extended I-component and the D-component. The I-component associated with LH9 is already ionized mostly (Figure \ref{fig9}c). {In the previous expansion-shell trigger, the first OB association was given as the initial condition.} This scenario explains the formation of the first OB association and the other OB associations consistently by the collisional compression.}

 \begin{table*}
 \begin{center}
  \caption{Comparison of gas momentum and stellar wind momentum}
         \label{tab:mom}
         \begin{tabular}{cccccccc}
         \hline \hline
          \multirow{3}{*}{Object}&$M$H{\sc i}& $M_{\rm CO}$ & $M$H{\sc i} + $M_{\rm CO}$&Momentum\footnotemark[$\dagger$]&$\#$ of&$\#$ of &Momentum\footnotemark[$\dagger$$\dagger$]\\
          &(I + D)&(I + D)&  &of gas & O-stars&WR stars& of stellar wind\\ 
          &{[}10$^6 M_{\odot}${]}&{[}10$^5 M_{\odot}${]}&{[}10$^6 M_{\odot}${]}&{[}10$^5 M_{\odot}$ km s$^{-1}${]}&&&{[}10$^5 M_{\odot}$ km s$^{-1}${]}\\ 
        (1)&(2)&(3)&(4)&(5)&(6)&(7)&(8)\\ \hline
        N11& 2.3& 4.5& 2.8&840&67&2&1.4  \\
        N79&2.6 & 3.7& 3.0&900 &17 &2&0.4  \\ \hline
       \end{tabular}
    \end{center}
    \begin{tabnote}
  Column (1): Object name. Column (2): H{\sc i} mass of the D- and I-component within a radius of 200 pc. Column (3): CO mass of the D- and I-component within the radius of 200 pc. Column (4): Total mass of H{\sc i} and CO. Column (5):  Momentum of gas. Column (6): Number of O stars. Column (7): Number of WR stars. Column (8): Total momentum of stellar wind of O/WR stars.\\
 \footnotemark[$\dagger$] We assumed that expansion velocity of gas is 30 km s$^{-1}$, which is velocity difference between the L- and I-components.
 \footnotemark[$\dagger$$\dagger$] We used momentum of stellar wind of O/WR stars calculated by mass loss rate and wind velocity (Abbott et al. 1982; Kudritzki \& Puls 2000). $D_{\rm mom}$ = $m_{\rm loss}$$\times$$v$$\times$$t$($m_{\rm loss}$: mass loss rate, $v$: wind velocity, $t$: wind duration time). O stars: $D_{\rm mom}$$\sim$2000 [$M_{\odot}$ km s$^{-1}$] ($m_{\rm loss}$$\sim$10$^{-6}$$M_{\odot}$ yr$^{-1}$; $v$$\sim$2000 km s$^{-1}$; $t$$\sim$10$^{6}$ yr). WR stars: $D_{\rm mom}$$\sim$2500 [$M_{\odot}$ km s$^{-1}$] ($m_{\rm loss}$$\sim$3$\times$10$^{-5}$$M_{\odot}$ yr$^{-1}$; $v$$\sim$2500 km s$^{-1}$; $t$$\sim$3$\times$10$^{5}$ yr).
   \end{tabnote}
   \end{table*}

In N11, the velocity separation of {the D-component and the I-component} is about 20--30 km s$^{-1}$ as shown in Figure \ref{fig9}(c). {Considering} a projection effect, the actual velocity separation is {more significant} than observed. This value is roughly consistent with the typical velocity separation in the high-mass star-forming regions triggered by colliding H{\sc i} flows in the LMC and M33: e.g., R136 ($\sim$50--60 km s$^{-1}$; Paper I), N44 (30--60 km s$^{-1}$; Paper II), and NGC 604 ($\sim$20 km s$^{-1}$; \cite{2018PASJ...70S..52T}). These values are also consistent with the predicted relative velocities of colliding clouds (10--60 km s$^{-1}$) in the LMC shown by \citet{2004ApJ...602..730B}.

{The shell expansion picture based on the ionized gas} estimated the total kinetic energy supplied by the high mass stars to be 2.3$\times$10$^{50}$ erg (Meaburn et al. 1980; Rosado et al. 1996). The present study revealed that the H{\sc i} gas is dominant in N11, and has a kinetic energy larger than 10$^{51}$ erg for the mass and velocity of the I-component. 
Considering that only part of the stellar energy can be converted into the H{\sc i} kinetic energy due to a small coupling efficiency {of $\sim$20\% (e.g., Weaver et al. 1977)}, the stellar feedback is not important in driving the gas. The tidally driven gas motion is a plausible picture of the {origin of the} I-component origin.}

{A difference from the H{\sc i} Ridge region is that the tidally driven L-component in N11 is much weaker than in the H{\sc i} Ridge (Figure \ref{fig2}a). 
We suggest that the I-component is mainly triggering the high-mass star formation in N11 at a smaller collision velocity than in the H{\sc i} Ridge region. 
This is consistent with the nearly solar dust-to-gas ratio derived from the $Planck$ dust optical depth (Tsuge et al. 2024 in prep.) and $A_{\rm v}$/ {$N$H ratio of the L- and D-components is 1/2 of the Galactic value, suggesting there is no strong inflow of the low-metal gas due to tidal interaction (Furuta et al. 2022).}}

\subsubsection{{N77-N79-N83 complex} region}\label{sec:4.1.4}
N79 is a H{\sc ii} region cataloged by \citet{1956ApJS....2..315H} located in the southwestern corner of the LMC. Figure \ref{fig10}a shows the H$\alpha$ image of the region (MCELS; Smith \& MCELS Team 1999). {The three star-forming regions, {N83}, N79, and {N77}, are located with a relative separation of 500--700 pc from each other (Figure \ref{fig10}a). They are associated with the emission nebulae and GMCs.
Distributions of the I- and D-components are shown in Figure \ref{fig10}b. Figure \ref{fig10}c shows the coordinate used in the position-velocity diagram shown in Figure \ref{fig10}d. Figures \ref{fig10}e and \ref{fig10}f are close-up images of Figure \ref{fig10}b toward {N83} and N79, respectively. }

The H{\sc i} features are the I-component and D-component, while no L-component is seen (Figure \ref{fig10}b). The I-component is distributed over a large area and consists of two major features of 300--500 pc sizes elongated in the E-W direction. In addition, there are more than ten small structures of the I-component. The D-component has high H{\sc i} intensity around the three regions and tens of small features of the D-component are distributed over a kpc scale. {We find that a filamentary H{\sc i} feature consisting of the I- and D-components appears to be elongated between {N83} and N79, {shown by red dashed lines of Figure \ref{fig10}b}. This H{\sc i} filament is $\sim$250 pc long with a width of $\sim$50 pc. We also find that the I-component is overlapped with the D-component toward {N77} etc. Figure \ref{fig10}d shows a position-velocity diagram taken along the H{\sc i} filament {in the coordinates for the grey Zone in Figure \ref{fig10}c}, which indicates that the I-component is merged with the D-component toward the filament, exhibiting broad H{\sc i} emission of a $\sim$45 km s$^{-1}$ span in velocity{ toward y=-0.5d, -0.3d- -0.0d, 0.2 d.} Figure \ref{fig10}e{, a close-up view toward {N83},} shows that the I-component has a cavity toward {N83} {(blue dashed line of Figure \ref{fig10}e)}, and a GMC in the D-component is in the west of N{83}. {Suppose} we assume that the ionization evacuated the cavity, the northwestern edge of the I-component {before} the ionization coincides with the southeastern edge of the D-component, where the GMC is located. Figure \ref{fig10}f shows that N79 with many O stars is distributed toward the lower intensity I-component along with the GMC in the I-component in the south of N79.}

{Based on Figures \ref{fig10}b,d,e, and f, we suggest that the I-component and the D-component are colliding with each to form the H{\sc i} filament with enhanced density and velocity dispersion over a few 100 pc. The collision is supported by the bridge features as well as the enhanced velocity {dispersion} to about 20 km s$^{-1}$ toward the filaments {of N79 and N83} (Figure \ref{fig10}d). It is possible that this compression achieved a high H{\sc i} column density of $\sim$4$\times$10$^{21}$ cm$^{-2}$, leading to the formation of N{83} and N79. The location of the filament coincides with the northwestern edge of the I-component, which shows a complementary distribution with the D-component (Figure \ref{fig10}b). 
We follow the same argument with N11 and adopt a view that the tidally-driven H{\sc i} flows play a role in triggering star formation. N83 and N79 seem to be younger than N79W as inferred from the higher brightness of the H$\alpha$ emission. 
N77 is the most evolved OB association, where the stellar feedback may have dispersed the H{\sc i} gas and reduced the velocity span.} {It seems to be still possible that N77 was also triggered by the collision between the I- and D-components at an epoch earlier than in N83 and N79.}

{We find that a collision similar to N11 is observed in a Galactic H{\sc ii} region Sh2-233 \citep{2022MNRAS.515.1012Y}. In Sh2-233, two CO clouds are colliding at their edges which are nearly straight, and a filamentary CO cloud {with} dense cloud cores is formed in the interface between the two clouds. Some of the dense cores in the filamentary cloud are massive enough to form high-mass protostars, and the densest core among them is suggested to be a typical, very young protostar IRAS 05358+3543 with {an} outflow, which attracts {the} keen interest of the community (e.g., \cite{2007A&A...466.1065B}). The Sh2-233 cloud is therefore an edge-on collision case similar to N11. It may also be relevant that the Musca CO filament show a similar circumstance where two H{\sc i} flows {collide} to cause the filamentary formation, while no star formation is seen yet \citep{2020A&A...644A..27B}.}

{In the region,} the most luminous YSO (H72.97- 69.39) in the LMC is found toward N79 \citep{2017NatAs...1..784O}. It was proposed that H72.97-69.39 might be a candidate for a superstar cluster because of its high star formation efficiency {(the fraction of mass that is transformed into stars) per free-fall time is 0.27--0.75 (Ochsendorf et al. 2017).} Subsequently, \cite{2019ApJ...877..135N} revealed filamentary CO clouds associated with the SSC candidate. Other mechanisms were suggested to be working in active star formation such as accretion flows \citep{2015Natur.519..331T} or tidal interactions (Bekki \& Chiba 2007a). The present scenario is along the tidal interaction picture and supports that the high-mass star formation is triggered by tidally-driven colliding H{\sc i} flows in the {N77-N79-N83 complex}.

\subsection{{{Comparison between regions}}}\label{sec:4.2}

{The properties of the gas and high mass stars are summarized for the four regions, the H{\sc i} Ridge, N44, N11, and {N77-N79-N83 complex}, in Table \ref{tab:comparison1}, where maximum $N$H{\sc i}, number of high mass stars, stellar age, size of the H{\sc i} collision, and mass of the relevant component are listed along with GMC mass, collision velocity and collision timescale. It is obvious that $N$H{\sc i}, is the largest, 5.8$\times$10$^{21}$ cm$^{-2}$, in R136, the most active high mass cluster including the largest number of high mass stars of 110 stars within 50 pc. We note that this $N$H{\sc i} may be even larger significantly if the ionization by the cluster is considered. The typical $N$H{\sc i} {within 50 pc from H{\sc ii} region of N44, N11, N83, and N79 is 1.0$\times$10$^{21}$ cm$^{-2}$, 1.5$\times$10$^{21}$ cm$^{-2}$, 0.8$\times$10$^{21}$ cm$^{-2}$, and 1.6$\times$10$^{21}$, respectively. }
{We also calculated the external pressure of the I-component $P_{e}$ from the cloud mass, size, and colliding velocity following the equation; $P_{\rm e}$= $\frac{3\Pi \it M v^{2}}{4 \pi R^{3}}$ = $\rho_{e} v^{2}$ (Elmegreen 1989). $M$ is the cloud mass, $v$ is the colliding velocity (difference of peak velocity of colliding clouds), $R$ is the radius of the cloud, and $\rho_{e}$ is the density of the cloud. $\Pi$ is defined by $\rho_{e}$=$\Pi$$\rho$, where $\rho$ is the mean density in the cloud. We adopt $\Pi$ = 0.5 (\cite{2015ApJ...806...35J,2019ApJ...874..120F}). We calculated the pressure of the I-component using physical properties within 50 pc from the H{\sc ii} region summarized in Table\ref{tab:comparison1}. There is a positive correlation between the external pressure and the number of O/WR stars. {The trend described aligns with the previous theoretical studies on YMC formation (e.g., Elmegreen \& Efremov 1997) and observational results of the Antennae galaxies (Johnson et al. 2015; Finn et al. 2019; Tsuge et al. 2021a; Tsuge et al. 2021b), which have shown a positive correlation between cluster mass and pressure. Thus, the high NHI, achieved by the strong compression due to the colliding H{\sc i} flow, maybe a key factor in determining the mass of a high-mass star cluster, whereas more details must await a full understanding of the gas-related over a large dynamic range from kpc to down to sub-pc. }


The proposed scenario of the H{\sc i} colliding flows that triggered the formation of high-mass stars has a wealth of outputs that are testable by future observations. One is high-resolution studies of the molecular gas with ALMA. For one of the molecular clouds of {N77-N79-N83 complex}, filamentary molecular clouds have already been found with ALMA \citep{2019ApJ...877..135N}, which is consistent with a collision formed filamentary clouds {scenario}. Follow-up observations of filamentary clouds toward {the} N44, N11, and {N77-N79-N83 complex} by using ALMA (2019.2.00072.S, 2021.1.00490.S, {PI: Tsuge, K}) are in progress. Further, observations of the H{\sc i} gas at a resolution comparable to ALMA observations will allow us to understand better the physical connection between filaments and large-scale H{\sc i} flows. Another direction is to pursue the metallicity in the gas in the LMC. The L- component is metal-poor toward the H{\sc i} Ridge and N44 in the NW L-component as shown in Paper I/II. Bekki \& Chiba (2007b) also found that the metal-poor gas from the SMC has continued to flow in the LMC since 0.2 Gyr ago (see also Figure 1 of Bekki \& Chiba 2007a). Further, there is increasing evidence for the low metallicity in the H{\sc i} Ridge, as opposed to the claim by Nidever et al. (2008). First, Olsen et al. (2011) measured the metallicity of stars in the H{\sc i} Ridge and showed that the low metallicity there is consistent with the metallicity of the SMC gas. Accordingly, Olsen et al. concluded that the gas in the H{\sc i} Ridge is coming from the SMC. Fukui et al. (2017) derived the dust-to-gas ratio in the H{\sc i} Ridge by using the 353 GHz dust emission and derived the ratio consistent with $Z$/$Z_{\odot}$=0.5 of the LMC disk. This is further supported by an independent study of $A_{\rm v}$ toward stars by Furuta et al. (2019; 2021; 2022). {In a separate paper,} we will present a detailed Metallicity distribution over the whole LMC elsewhere.

\subsubsection{{The alternative interpretation}}
{We note that there is another interpretation that most of the gas in the Magellanic System was stripped off from the LMC but not from the SMC, as claimed by Nidever et al. (2008). These authors referred to the H{\sc i} super giant shells (=SGSs) identified by Kim et al. (1999) and claimed that the L- and D-components, H{\sc i} Arms E and B (Staveley-Smith et al. et al. 2003) which appear to be extensions of the $''$SEHO$''$ (=$"$southeastern H{\sc i} overdense$''$)=$''$the H{\sc i} Ridge$''$ in the present paper, as well as the Magellanic Stream (MS) and the Leading Arm Feature (LAF) were created by the action of the SGSs, and estimated that in total $\sim$20,000 SNe’s are required to accelerate and create the MS and LAF. These authors claimed that the number of SNe’s is explicable by the present rate of SNe’s. The acceleration claimed by Nidever et al. (2008) contradicts the above papers, and we examine the difference between the two interpretations;}

{If we assume that SGSs are formed in the H{\sc i} Ridge, the mass of the “SGSs” can be estimated to be $\sim$$>$3$\times$10$^7$ $M_{\odot}$ from the PV diagram (Figure \ref{fig21}b). For {an} expansion velocity of 50 km s$^{-1}$, the kinetic energy required to form the SGSs becomes $>$3$\times$10$^{53}$ erg. This energy can be supplied by 300 SNe if the energy is completely converted to the gas motion. However, the conversion efficiency of the SNe into gas kinetic energy is estimated theoretically to be $<$20 \% by theoretical numerical simulations (e.g., Weaver et al. 1987; Tomisaka et al. 2001). The efficiency is also estimated to be $<$10 \% via observations of H{\sc i} super shells which are likely driven by 20--30 SNe (e.g., Fukui et al. 2001; Suad et al. 2019). The efficiency lower than 100 \% is due to the radiative energy loss and the escaping momentum through an inhomogeneous H{\sc i} shell. This means that at least 300$\times$(5--10)=1500--3000 SNe are required to create the SGSs having a kinetic energy of $>$3$\times$10$^{53}$ erg.

Since the number of O-stars in the present SGSs is $\sim$400 within 100 pc of R136, we infer that the high-mass stars cannot create the SGSs energetically in the H{\sc i} Ridge. Considering that these O/WR stars are the most active high mass cluster in the Local Group, we presume that the MS and LAF, even more massive than the H{\sc i} Ridge (4.1$\times$10$^8$ $M_{\odot}$), cannot be explained by the SNe energy either, unless we assume most luminous clusters are continuously formed in the H{\sc i} Ridge.

{In Figure \ref{fig20} of Appendix 3, we show that these O/WR stars uncorrelated with the I component are found to be located inside the supergiant shells (SGSs) catalogued {at least in three of them, SGS3, SGS4, and SGS7, which include nearly 50 O/WR stars}\citep{1976MmRAS..81...89D,1980MNRAS.192..365M,1990ApJ...365..510C,1999AJ....118.2797K,2013ApJ...763...56D}. This raises a concern on the role of SGSs in evacuating the H{\sc i} gas. If we assume that the stellar feedback evacuates the I-component via ionization/stellar winds soon after the star formation in SGSs within $\sim$10 Myr, the O/WR stars uncorrelated with the I-component can be explained.  }

\section{{Future prospects}}

{Based on the findings of the present study, we will proceed with multi-wavelength astronomy to reveal the processes of high-mass star formation and the evolution of interstellar gas in the galaxies. In an upcoming paper, we intend to investigate the formation mechanisms of molecular clouds using the ALMA CO project accepted for ALMA Cycle 8 2021 (project code 2019.2.00072.S and 2021.1.00490.S, PI: K. Tsuge). We plan to report on the observational results of molecular cloud data from N44, N11, and the N77-N79-N83 complex. Additionally, we plan to validate the outflow and inflow of gas across the entirety of the LMC through a comparison of the spatial distribution of dust-to-gas ratio, as investigated by studies such as Fukui et al. (2017) and Tsuge et al. (2019) with the dynamics of H{\sc i} (Tsuge et al., 2024 in press.) due to tidal interactions. Furthermore, we can study the heating processes of diffuse X-ray emissions and the 3D structure of collisional gas by incorporating the latest X-ray data obtained by the extended Roentgen Survey with an Imaging Telescope Array (eROSITA; Merloni et al. 2012; Predehl et al. 2021) . We proposed the heating scenario of the diffuse X-ray emissions due to H{\sc i} gas collisions and investigated the 3D structure of collisions toward the H{\sc i} Ridge region (Knies et al., 2021). The comparison of HI, CO, and X-ray results in the N11 region has also been submitted (Tsuge et al., 2024). We plan to expand such multi-wavelength analyses to cover the entire LMC.}

\section{Conclusions}\label{Conclusions}
{In order to explore the role of the H{\sc i} colliding flows on high-mass star formation in the LMC, }{we {have} comprehensive{ly} analy{zed} of the H{\sc i} data over the whole LMC} {{and confirmed} with the new numerical simulations of the gas driven by the tidal interaction.} The main conclusions of the present paper are summarized as follows;

\begin{enumerate}
\item   We analyzed the H{\sc i} data at {a} resolution of 60$\arcsec$ corresponding to $\sim$15 pc \citep{2003ApJS..148..473K}. The spatial distribution of the I-component {was} revealed over the whole LMC for the first time. The I-component is the intermediate velocity component between the two H{\sc i} components (the L- and D-components) with {a} velocity difference of $\sim$50 km s$^{-1}$. {We interpret that the I-component was produced by velocity shifts of the two colliding clouds, i.e., the L- and D-components, due to the collisional interaction.}

\item The distribution of the I-component exhibit{s} spatial correlation {with} the high-mass {(O/WR)} stars over the whole LMC, and 74 \% of the high-mass stars are associated with the I-component. {This trend significantly differ{s} from the purely random distribution and lends support for that the high-mass stars are physically connected to the I-component. We interpret that the gas compression driven by the H{\sc i} collisions triggered the formation of molecular clouds and high-mass stars.}

\item  {In particular,} we revealed the detailed spatial and velocity distributions of the {H{\sc i} gas toward the three outstanding high-mass star forming regions, i.e., the} H{\sc i} Ridge region in the southeast of the LMC, the N11 region and the {N77-N79-N83 complex} region {in the western Arm} with a velocity difference of $\sim$10--60 km s$^{-1}$. The collisions {are} characterized by the spatial complementary distribution (anti-correlated distribution) and bridge features in velocity space{, which are typical signatures of a cloud-cloud collision}.

\item As the most outstanding case of the collision, we explored the geometry of the collision in the H{\sc i} Ridge. We found that the L-component there is probably plane-like with some tilt relative to the D-component in the sense that the northwestern part is closer to the D-component than the southeastern part. The tilt naturally explains the different epoch of the collision from place to place, thereby offering an explanation {of} the cause of the age difference among the high-mass stars in the H{\sc i} Ridge. {This picture offers the cause of no high-mass star formation in the southern half of the H{\sc i} Ridge, where the H{\sc i} collisions will happen soon with $\sim$10 Myr.} 

\item {We extended the {observational} analysis to the western Arm, including {the} N11 and {N77-N79-N83 complex}, and found that the tidal interaction is probably responsible for {forming} the I-component in the region. A difference from the H{\sc i} Ridge is that there is a little hint of the L-component, while the I-component is likely induced by the tidal interaction {likely induces the I-component}. The H{\sc i} gas shows signatures of collisions in the two regions, and we suggest that the H{\sc i} colliding flows consisting of the I- and D-components are triggering high-mass star formation in this part of the galaxy at a lower speed than in the H{\sc i} Ridge.}

\item {A comparison with R136, N44, N11, and {N77-N79-N83 complex} suggests that the sequence of the number of high-mass stars could be understood to be due to density and collision velocity, which produce high ambient pressure.}

\item {Thanks to the unrivaled small distance of the LMC, the present paper has demonstrated, with an unprecedented detail better than 10--100 pc, that the H{\sc i} colliding flows driven by the galactic tidal interaction is likely a crucial mechanism {that} triggers high-mass star formation over the whole galaxy. This insight, which has been discussed often in the previous papers (see references in the introduction), has profound implications on the galaxy evolution in the {U}niverse, spanning over a Gyr, under the present findings at the high resolution of 10--100 pc. }

\end{enumerate}

\begin{ack}
We would like to thank the anonymous referees for their careful reading of the manuscript and constructive suggestions which helped to improve the manuscript. The NANTEN project is based on a mutual agreement between Nagoya University and the Carnegie Institution of Washington (CIW). We greatly appreciate the hospitality of all the staff members of the Las Campanas Observatory of CIW. We are thankful to many Japanese public donors and companies who contributed to the realization of the project. This study was financially supported by JSPS KAKENHI Grant Number 15H05694. This work was also financially supported by Career Development Project for Researchers of Allied Universities. The ATCA, Parkes, and Mopra radio telescope are part of the ATNF which is funded by the Australian Government for operation as a National Facility managed by CSIRO. The UNSW Digital Filter Bank used for the observations with the Mopra Telescope was provided with support from the Australian Research Council. Based on observations obtained with Planck, an ESA science mission with instruments and contributions directly funded by ESA Member States, NASA, and Canada. The Southern H-Alpha Sky Survey Atlas, which is supported by the National Science Foundation. Cerro Tololo Inter-American Observatory (CTIO) is operated by the Association of Universities for Research in Astronomy Inc. (AURA), under a cooperative agreement with the National Science Foundation (NSF) as part of the National Optical Astronomy Observatories (NOAO).  The MCELS is funded through the support of the Dean B. McLaughlin fund at the University of Michigan and through NSF grant 9540747. This work was supported by Grant-in-Aid for JSPS Fellows Number 23KJ0322.
\end{ack}

\renewcommand{\arraystretch}{1.0}
\begin{table*}[HBTP]
\centering

\rotatebox{90}{
\begin{minipage}{1.0\textwidth}
      
  \caption{Physical properties of collision}%
  \begin{adjustwidth}{-1in}{-1in} \begin{center} \resizebox{1.2\textwidth}{!}{ 
  \scalebox{1.0}{
  \begin{tabular}{cccccccccccccc}
    \hline \hline
\multirow{3}{*}{Object}& $N$H{\sc i} & $\#$ of & Age & Size of \footnotemark[$\dagger$] & $M$H{\sc i}\footnotemark[$\dagger$$\dagger$] &$M$H{\sc i}\footnotemark[$\dagger$$\dagger$]&$M$H{\sc i}\footnotemark[$\dagger$$\dagger$] &$M$$_{\rm CO}$&$n$ (I) &$P/k_{\rm B}$ (I)&$v_{\rm col}$\footnotemark[$\dagger$$\dagger$$\dagger$]&$t_{\rm col}$\\ 
    &(max) &high-mass &\  & H{\sc i}  collision& (L) &(I) &(D) &(total) &&10$^5$&\ &\ \\
    &{[}10$^{21}$ cm$^{-2}${]} &stars &{[}Myr{]}& {[}pc{]} &  {[}10$^{5}$$M_{\odot}${]} & {[}10$^{5}$$M_{\odot}${]}&{[}10$^{5}$$M_{\odot}${]}&{[}10$^{5}$$M_{\odot}${]}&{[}cm$^{-3}${]}&{[}K cm$^{-3}${]}&{[}km s$^{-1}${]}&{[}Myr{]}\\
   (1)&(2)&(3)&(4)&(5)&(6)&(7)&(8)&(9)&(10)&(11)&(12)&(13)\\ \hline
   H{\sc i} Ridge&7.4&$\sim$300&---&$\sim$2$\times$1 kpc&200&260&400&140&---&---&\multirow{3}{*}{50 (70)}&\  \\
   \multirow{2}{*}{R136}&7.3&$\sim$180&\multirow{2}{*}{1.5--4.7\footnotemark[a]}& $R$$<$200 & 3.4&19&17&2.7&---&---&&$\sim$3 \\
   &5.8& $\sim$110& & $R$$<$50 & 0.2&{1.4}&0.8&1.9&10.8$\pm$0.1&32.0$\pm$9.1&&(Paper I)  \\  \hline
   \multirow{3}{*}{N44}&5.3& $\sim$40 & \multirow{3}{*}{5--6\footnotemark[b]}  &$\sim$450$\times$300  &11  &32&100&16.0&---&---&\multirow{3}{*}{50 (70)}&\\
   &2.0& $\sim$40& &$R$$<$200&2.6&4.4&18&15.6&---&---&&5--10\\  
   &1.0& $\sim$30& &$R$$<$50&0.3&{0.6}&1.3&3.3&4.6$\pm$0.1&13.7$\pm$3.9&&(Paper II)\\ \hline
   \multirow{3}{*}{N11}& 4.8& $\sim$70 & $\sim$1(LH9)\footnotemark[c]  &$\sim$800$\times$800 &0.4  &13&87&9.5&---&---&\multirow{3}{*}{30 (40)}&\multirow{3}{*}{$\sim$3}\\
   &1.5&  $\sim$63& $\sim$3.5(LH10)& $R$$<$200& 0.1&5.8&15.6&5.9&---&---&& \\ 
   & 1.5& $\sim$25& & $R$$<$50& 0.09&{0.9}&0.5&---&6.9$\pm$0.1&10.5$\pm$4.2&& \\  \hline
   \multirow{2}{*}{N79S,W, E}&3.9& \multirow{2}{*}{$\sim$20} & \multirow{4}{*}{$<$0.5\footnotemark[d]} & $\sim$500$\times$800  &---  &12  &76&4.9&&&\multirow{4}{*}{30 (40)}&\multirow{4}{*}{---\footnotemark[$\dagger$$\dagger$$\dagger$$\dagger$] }\\
   &1.6&  & & $R$$<$200&---&3.8&17&3.8&&&&\   \\
   N79 S& 1.6& $\sim$8& & $R$$<$50& ---&{0.7}&0.7&1.3&5.4$\pm$0.1&2.9$\pm$2.0&&\   \\
 N79 E& 0.8& $\sim$6 & & $R$$<$50& ---&{0.9}&1.0&1.6&6.9$\pm$0.1&3.8$\pm$2.5&& \\ \hline
  \end{tabular} }
  } \end{center} \end{adjustwidth} 
 \label{tab:comparison1}
  \begin{tabnote}
  Column (1): Object name. Column (2): Peak column density of H{\sc i}. Column (3): Number of O/WR stars \citep{2009AJ....138.1003B}. Column (4): Age of star-forming region. Column (5): Size of H{\sc i} collision. Column (6), (7), and (8) are H{\sc i} mass of the L-, I-, and D-components, respectively. Column (9) Total mass of the molecular cloud. Column (10):  Density of the I-component. Column (11): Compressed gas pressure. Column (12) Colliding velocity. Column (13): Time scale of the collision.\\
 \footnotemark[$\dagger$] Physical properties were also calculated for two different radius, within 50 pc and 200 pc.  
 \footnotemark[$\dagger$$\dagger$] We calculated H{\sc i} mass of the region enclosed by a contour of WH{\sc i} = 300 K km s$^{-1}$. As for the L-component of N11, we used the region enclosed by a contour of WH{\sc i} = 150 K km s$^{-1}$ because integrated intensity was very weak. As for mass of the I-component within 50 pc, we corrected the effect of ionization. We replace WH{\sc i} toward the H{\sc ii} region by the highest value of WH{\sc i} (I) within 50 pc of the H{\sc ii} region.  
 \footnotemark[$\dagger$$\dagger$$\dagger$] Values inside parentheses are colliding velocity assumed angles relative to line of sight is 45 deg. 
  \footnotemark[$\dagger$$\dagger$$\dagger$$\dagger$] We could not estimate time scale because there is no displacement between the two clouds.
  \footnotemark[a]{Schneider et al. (2018)}
\footnotemark[b]{Will et al. (1997)}
\footnotemark[c]{e.g., Walborn et al. (1999)}
\footnotemark[d]{Ochsendorf et al. (2017); Nayak et al. (2019)}
  \end{tabnote}
\end{minipage}}
\end{table*}
\renewcommand{\arraystretch}{1}
\clearpage

\appendix

\section{Velocity ranges of the L-, I-, and D-components}
Figure \ref{fig16} shows Histogram of the number of pixel whose brightness temperature of H{\sc i} is greater than 30 K toward the northern part of H{\sc i} Ridge (R.A.=86.045975 deg.--88.766197 deg., Dec.=$-$69.122518 deg.--$-$70.003540 deg.). There are three velocity components in the histogram and they are corresponding to the L-, I-, and D-components

\begin{figure}[htbp]
\begin{center}
\includegraphics[width=\linewidth]{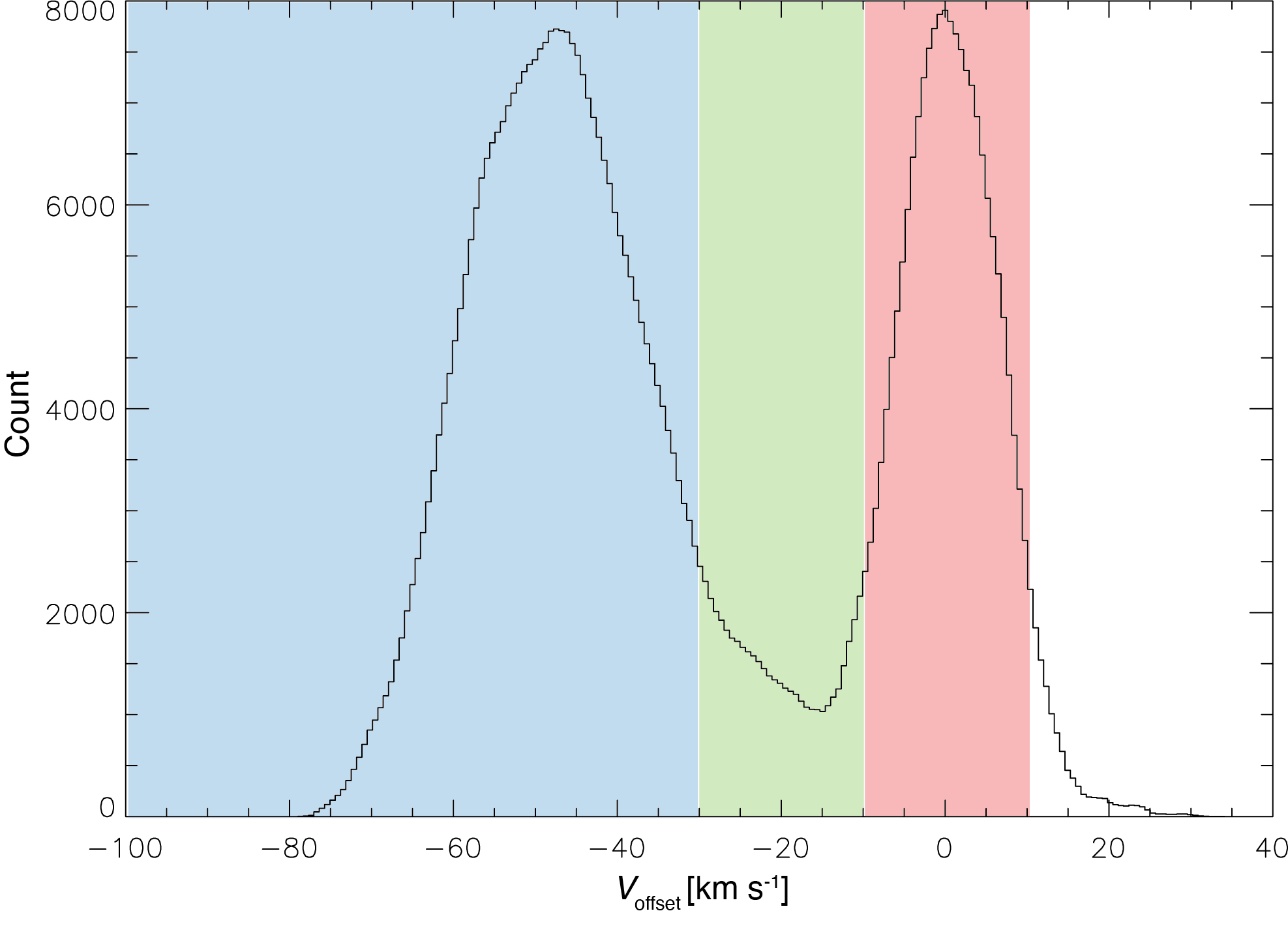}
\end{center}
\caption{{Histogram of the number of pixel whose brightness temperature of H{\sc i} is greater than 30 K toward the northern part of H{\sc i} Ridge (R.A.=86.045975 deg.--88.766197 deg., Dec.=$-$69.122518 deg.--$-$70.003540 deg.). The horizontal and vertical axis are $V_{\rm offset}$ and the number of pixel within the each $V_{\rm offset}$ bins, respectively. The blue, green, and red show the velocity ranges of the L-, I-, and D-components, respectively.}}  
\label{fig16}
\end{figure}%

\section{Velocity channel maps of H{\sc i} toward R136, N11, and N79}
Figures \ref{fig17}, \ref{fig18}, and \ref{fig19} are velocity channel maps of H{\sc i} toward R136, N11, and N79, respectively. The  velocity ranges from $-$48.1 to 39.6 km s$^{-1}$, with an interval of 3.25 km s$^{-1}$.

\section{Distributions of high-mass stars correlated with the I-component}
Figure \ref{fig20} shows the distributions of high-mass stars correlated with the I-component in (a) and not correlated with the I-component in (b). 

\section{Channel maps of the position velocity diagrams over the whole LMC}
We show the 11 right ascension--velocity diagrams of H{\sc i} over the whole LMC in Figure \ref{fig21}. The integration range is 0.27 deg. ($\sim$235.6 pc), and the integration range is shifted from north to south in 0.27 deg. step.

\begin{figure}[htbp]
\begin{center}
\includegraphics[width=8.5cm]{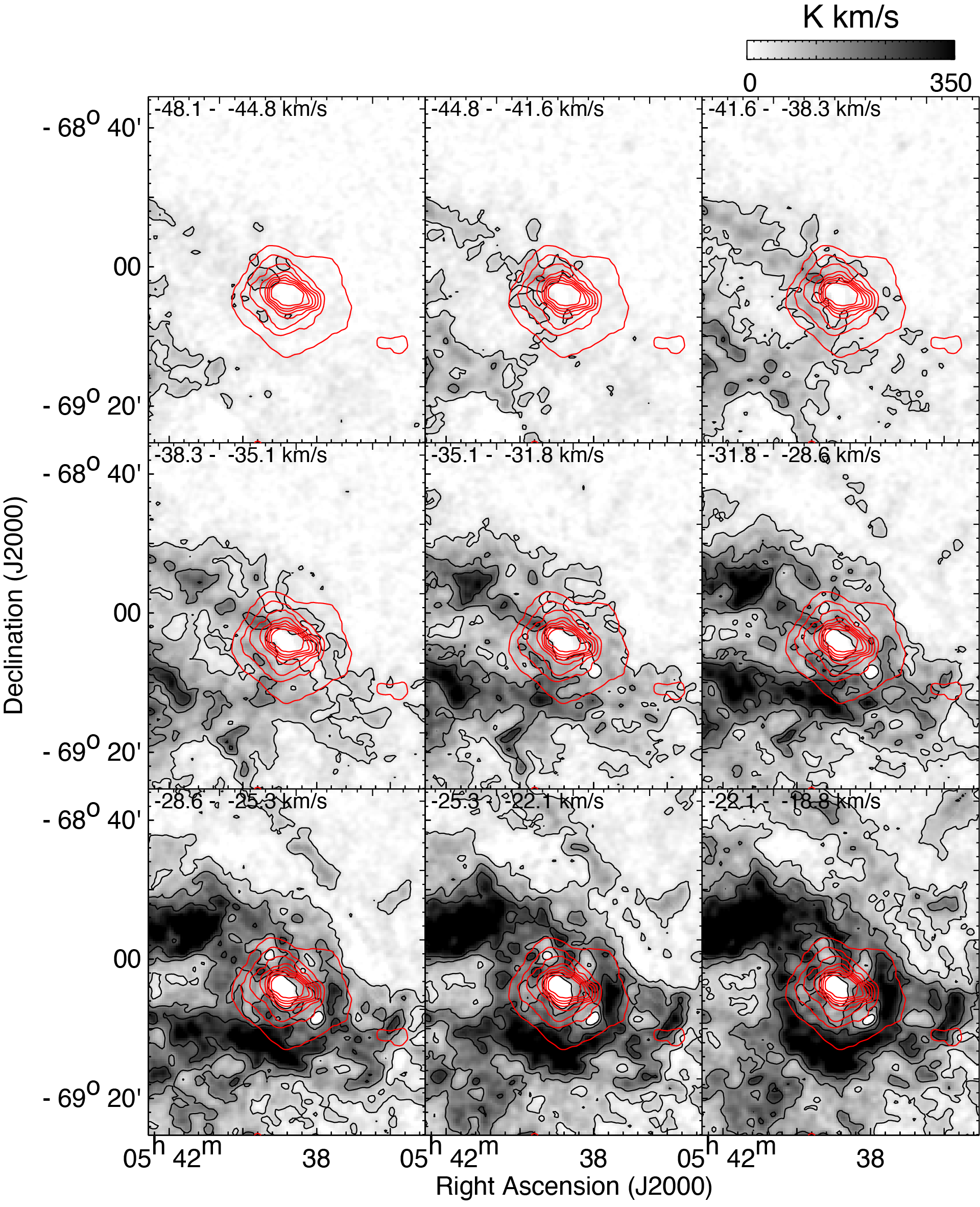}
\end{center}
\caption{Velocity channel maps of H{\sc i} gas with a velocity step of 3.25 km s$^{-1}$ overlaid with H$\alpha$ emission by {red} contours {toward R136}. The lowest level and intervals are 50 K km s$^{-1}$ and 100 K km s$^{-1}$ for H{\sc i} and 150 Rayleigh and 150 Rayleigh for H$\alpha$}  
\label{fig17}
\end{figure}%
\clearpage
\setcounter{figure}{13}
\begin{figure*}[HTBP]
\begin{center}
\includegraphics[width=8.5cm]{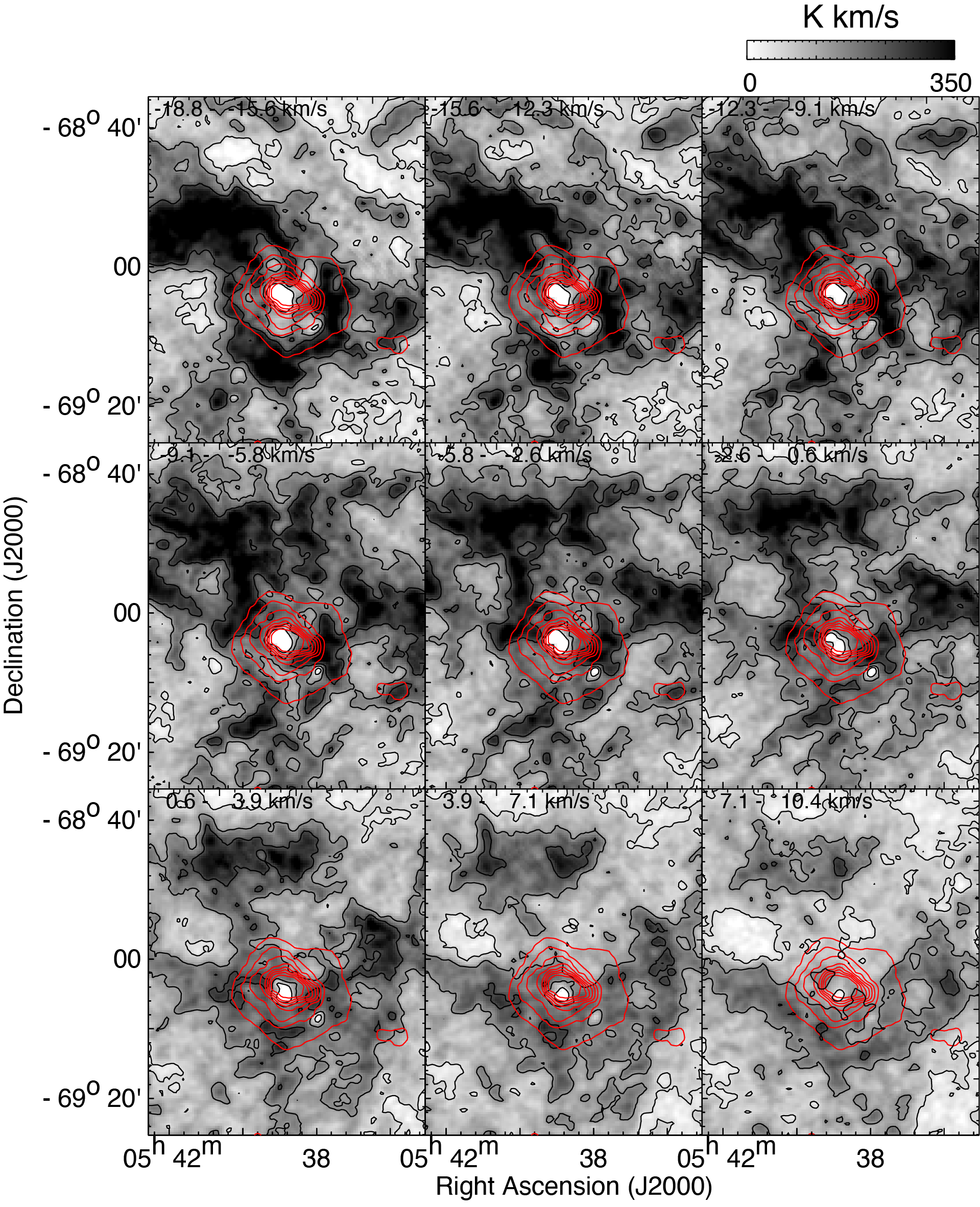}
\end{center}
\caption{Continued.}  
\label{fig17}
\end{figure*}%

\setcounter{figure}{13}
\begin{figure*}[htbp]
\begin{center}
\includegraphics[width=8.5cm]{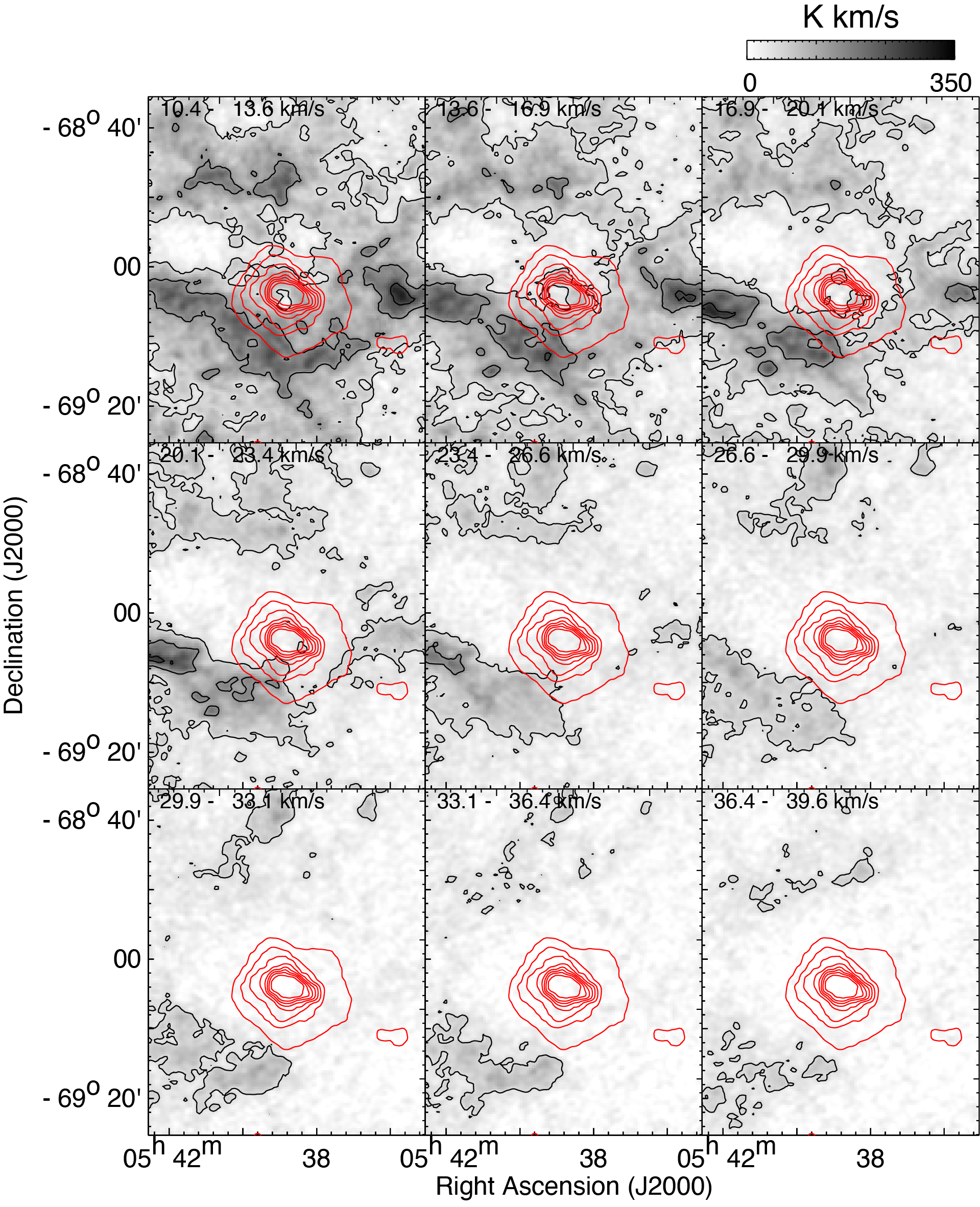}
\end{center}
\caption{Continued}  
\label{fig17}
\end{figure*}%

\begin{figure*}[htbp]
\begin{center}
\includegraphics[width=10cm]{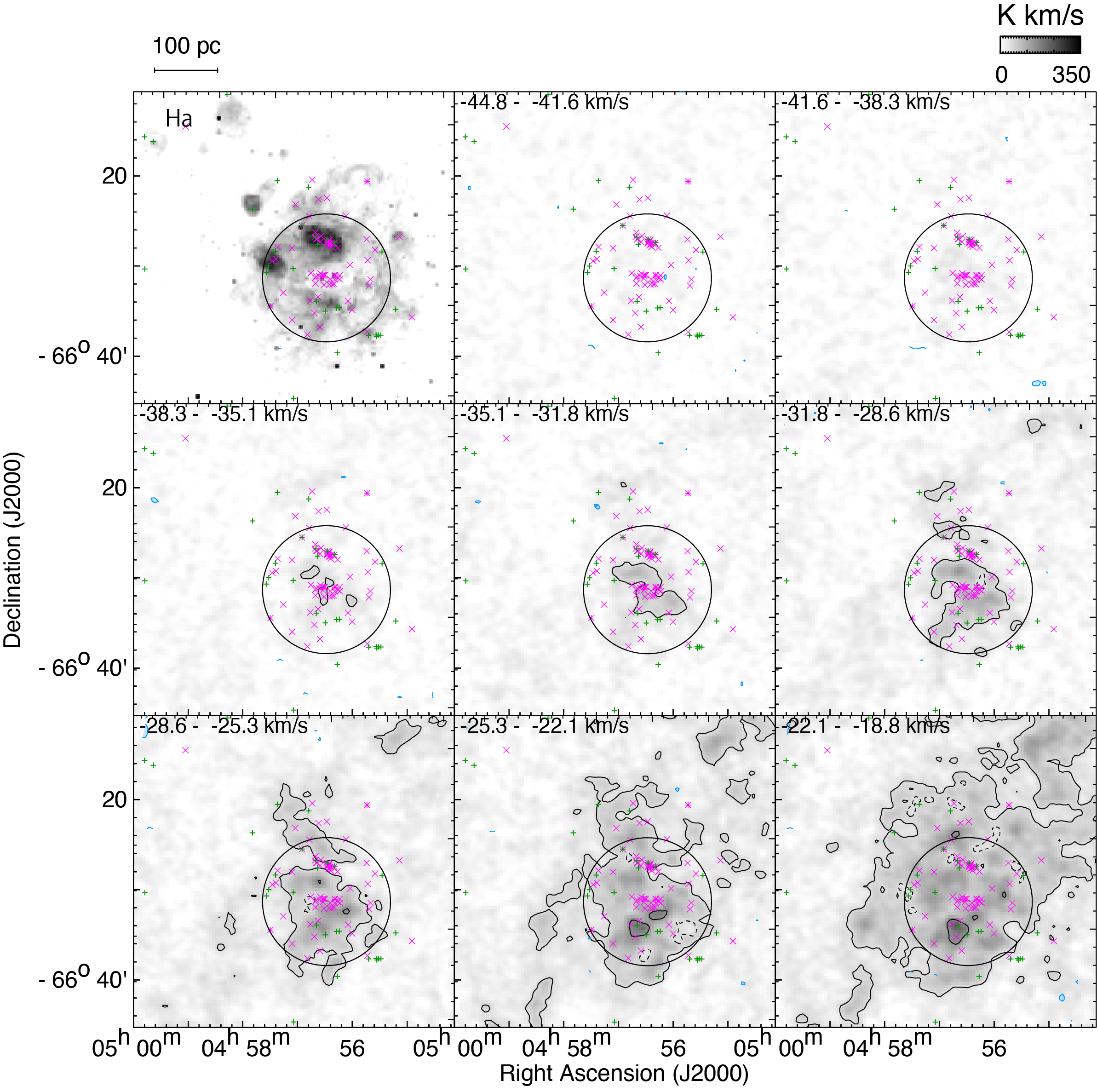}
\end{center}
\caption{Velocity channel maps of H{\sc i} gas with a velocity step of 3.25 km s$^{-1}$ {toward N11}. The lowest level and intervals are 50 K km s$^{-1}$ and 100 K km s$^{-1}$.  The black circle indicates a ring morphology with a cavity of $\sim$100 pc in radius, enclosing OB association LH9 (Lucke \& Hodge 1970). Magenta asterisks and crosses are WR-stars and O-type stars, respectively (Bonanos et al. 2009). The green crosses indicate young stellar objects cataloged by Seale et al. (2009).}  
\label{fig18}
\end{figure*}%

\setcounter{figure}{14}
\begin{figure*}[htbp]
\begin{center}
\includegraphics[width=10cm]{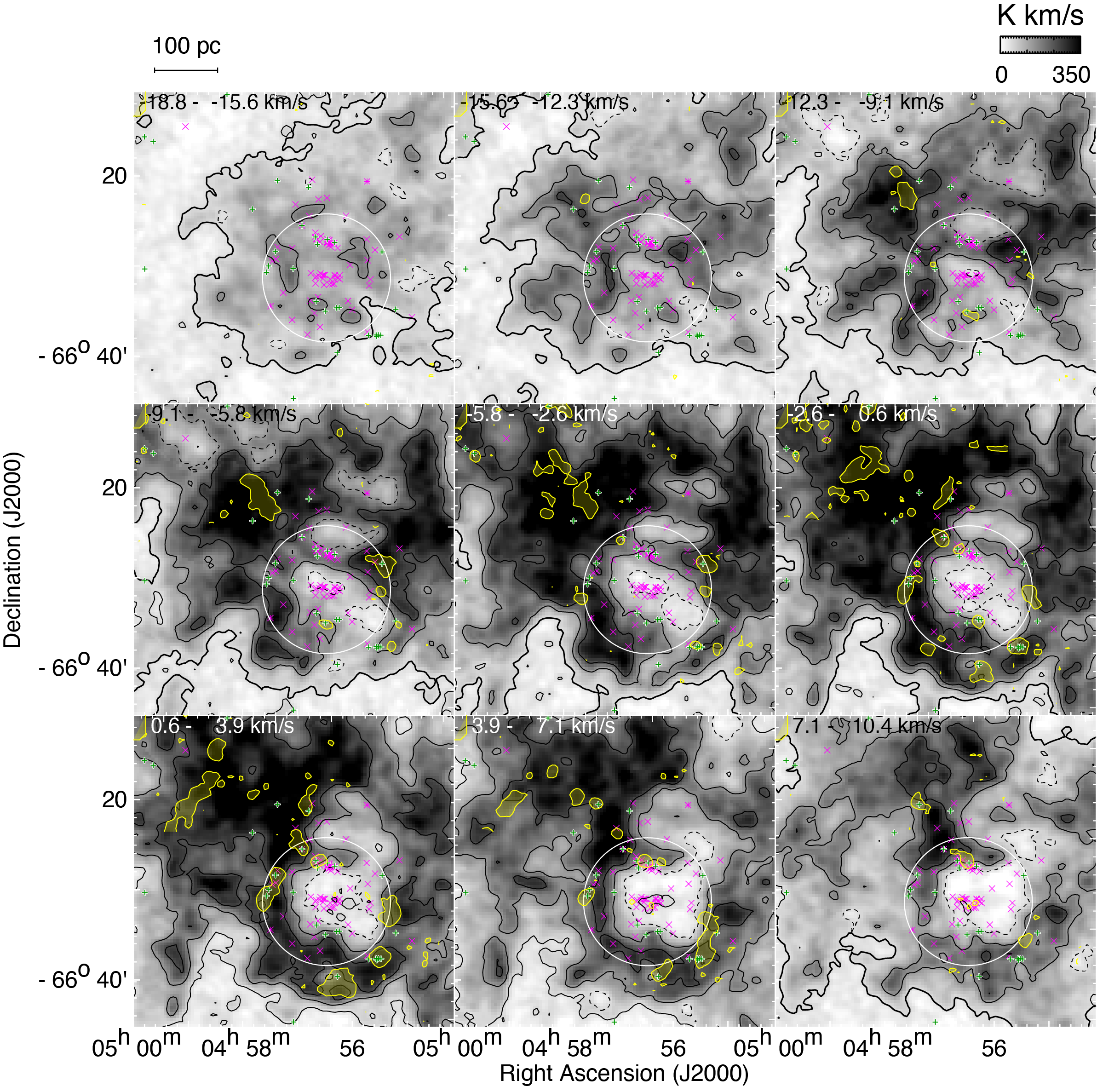}
\end{center}
\caption{Continued. The {yellow} contours indicate velocity channel maps of $^{12}$CO ($J$=1--0) obtained with Mopra telescope (Wong et al. 2011). The contour levels is $\sim$1.2 K km s$^{-1}$ (3 $\sigma$). }  
\label{fig18}
\end{figure*}%


\setcounter{figure}{14}
\begin{figure*}[htbp]
\begin{center}
\includegraphics[width=10cm]{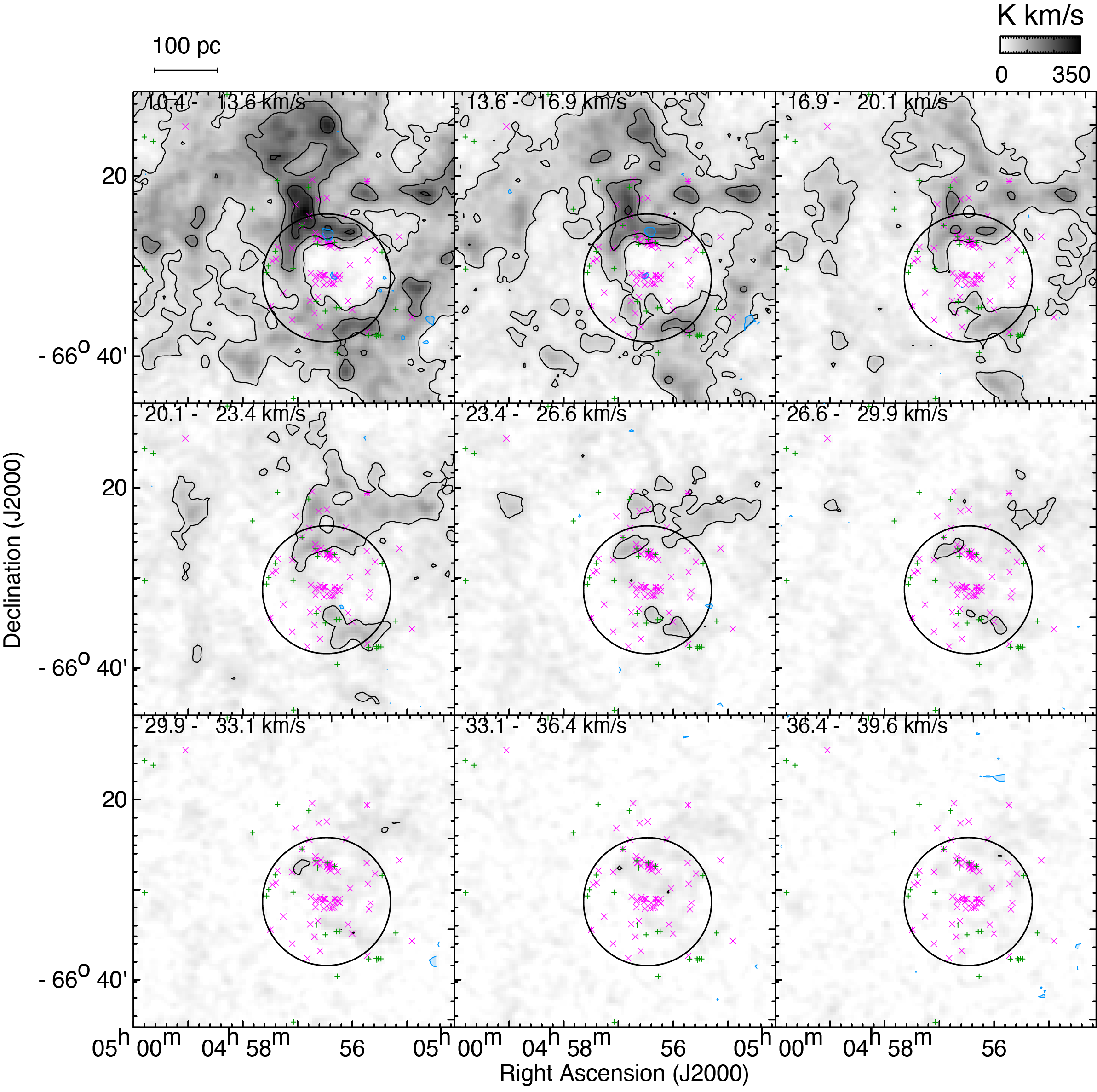}
\end{center}
\caption{Continued}  
\label{fig18}
\end{figure*}%

\begin{figure*}[htbp]
\begin{center}
\includegraphics[width=9.5cm]{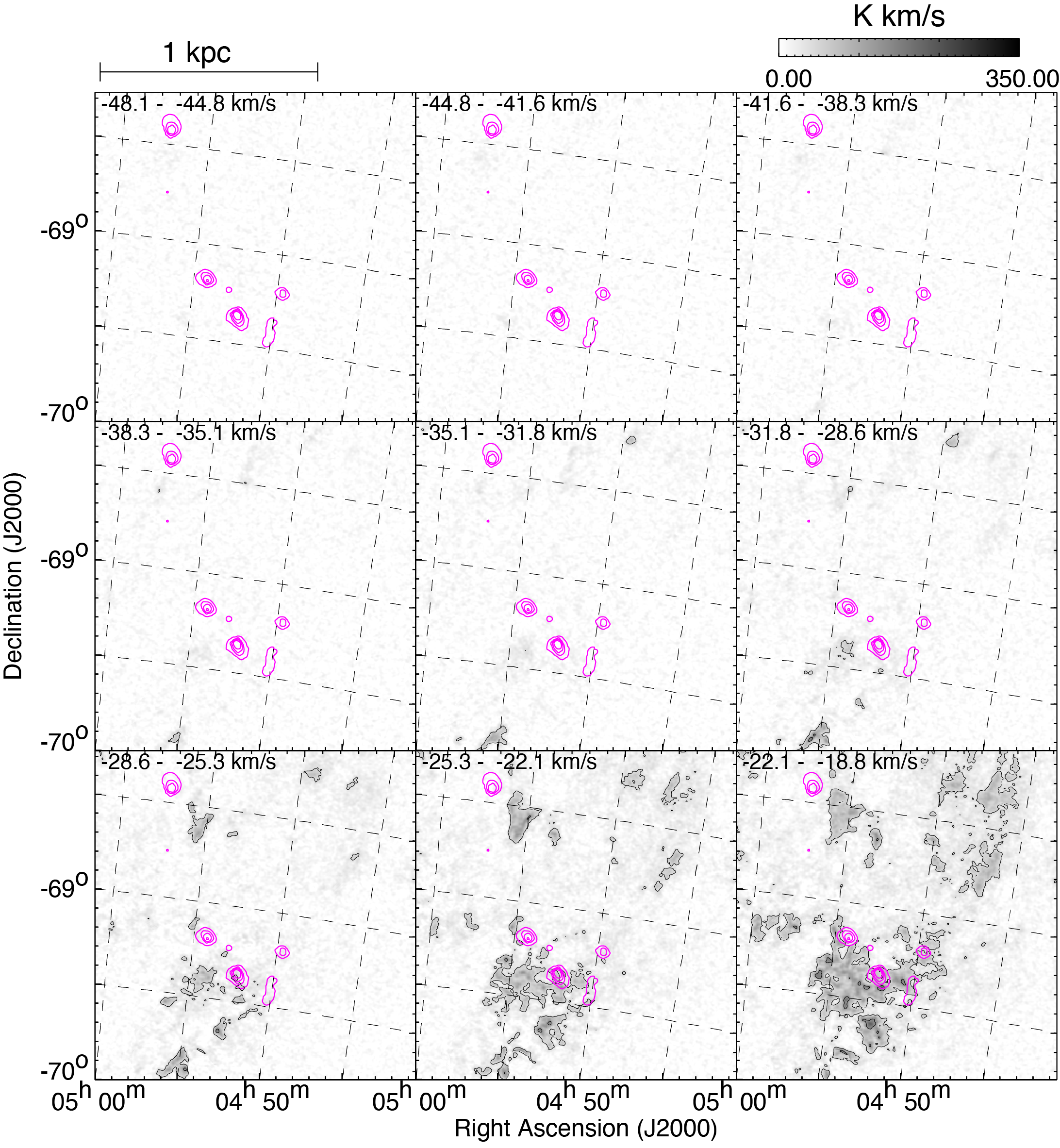}
\end{center}
\caption{Velocity channel maps of H{\sc i} gas with a velocity step of 3.25 km s$^{-1}$ overlaid with H$\alpha$ emission by {red} contours {toward N77-N79-N83 complex region}. The lowest level and intervals are 50 K km s$^{-1}$ and 100 K km s$^{-1}$ for H{\sc i} and 150 Rayleigh and 150 Rayleigh for H$\alpha$}  
\label{fig19}
\end{figure*}%

\setcounter{figure}{15}
\begin{figure*}[htbp]
\begin{center}
\includegraphics[width=9.5cm]{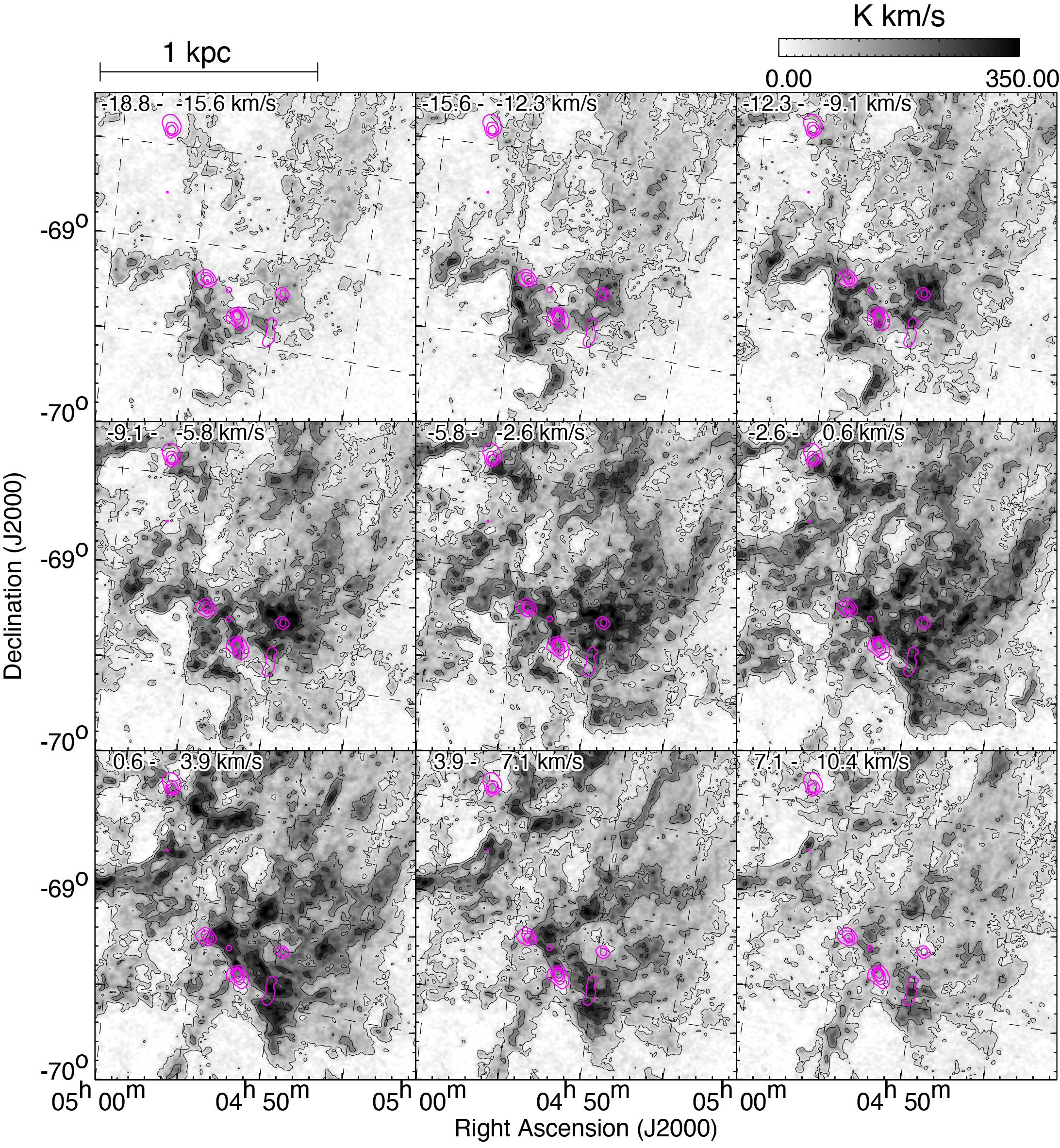}
\end{center}
\caption{Continued}  
\label{fig19}
\end{figure*}%

\setcounter{figure}{15}
\begin{figure*}[htbp]
\begin{center}
\includegraphics[width=9.5cm]{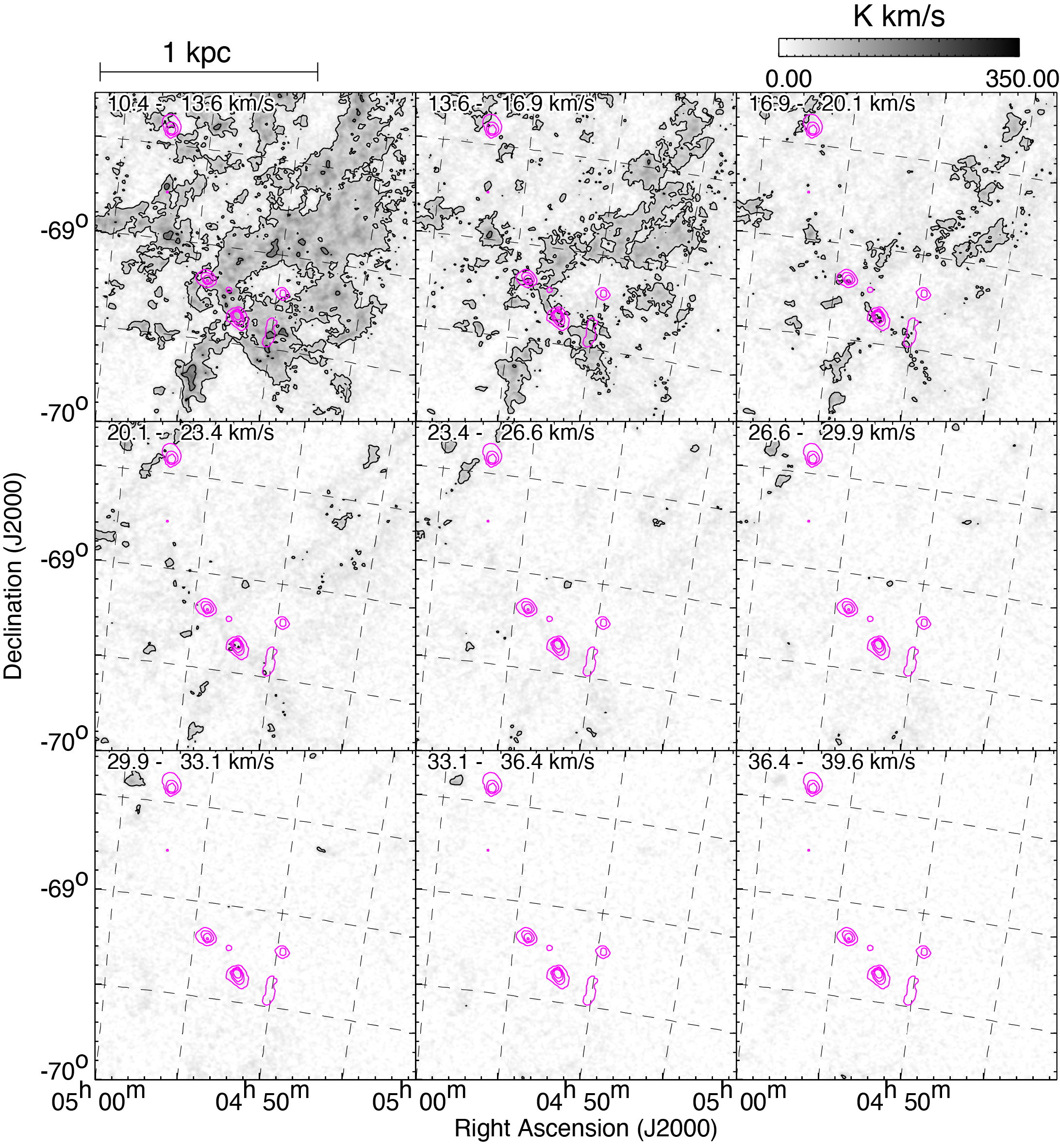}
\end{center}
\caption{Continued }  
\label{fig19}
\end{figure*}%

\setcounter{figure}{16}
\begin{figure*}[htbp]
\begin{center}
\includegraphics[width=\linewidth]{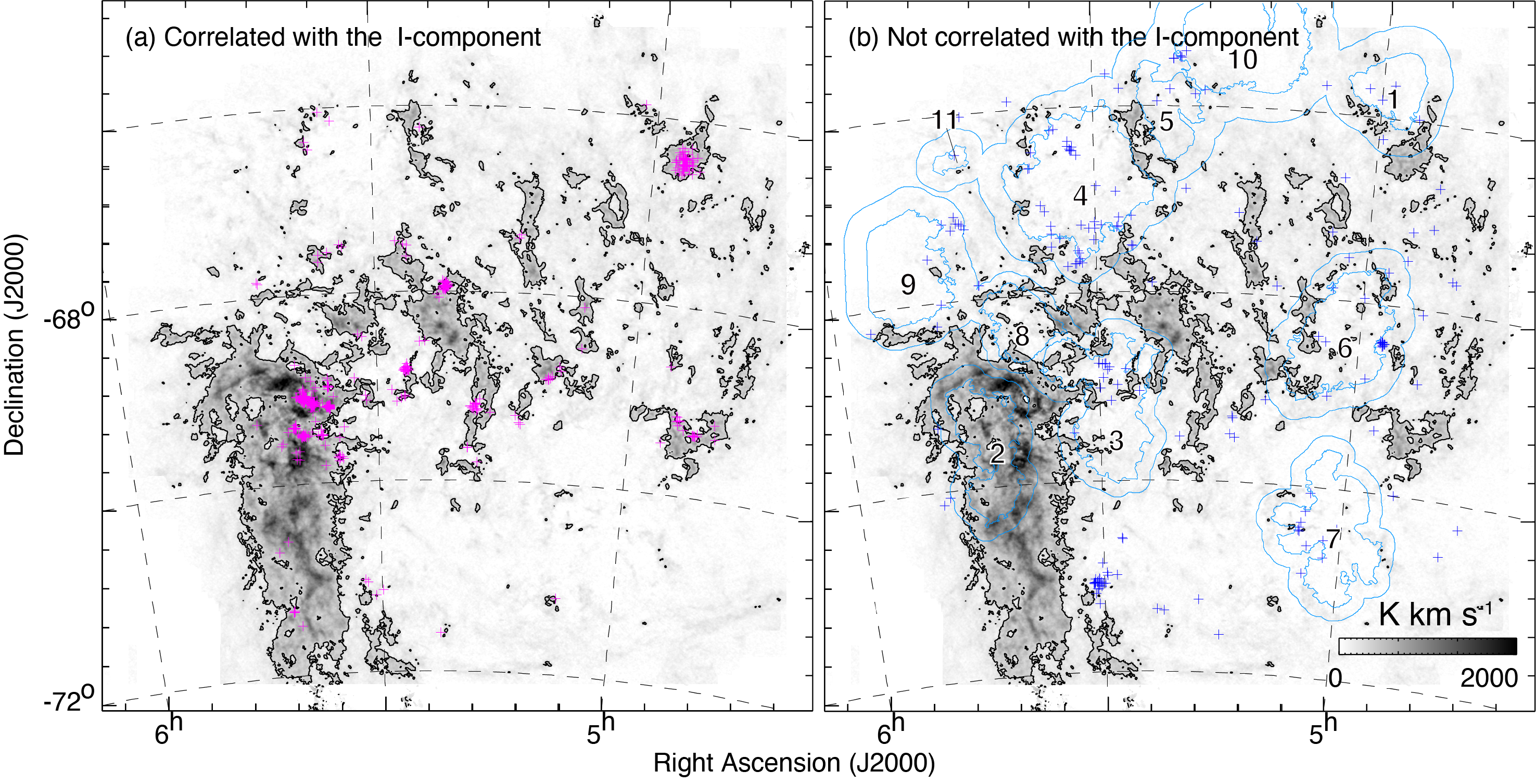}
\end{center}
\caption{(a) The distribution of high-mass stars correlated with the I-component by magenta crosses, which is corresponding to the blue histogram where integrated intensity is larger than 300 K km s$^{-1}$ in Figure 6(b). (b) The distribution of high-mass stars not correlated with the I-component by blue crosses. Light blue contours indicate the positions of super giant shells (Dawson et al. 2003). }  
\label{fig20}
\end{figure*}%

\setcounter{figure}{17}
\begin{figure*}[Htbp]
\begin{center}
\includegraphics[width=15cm]{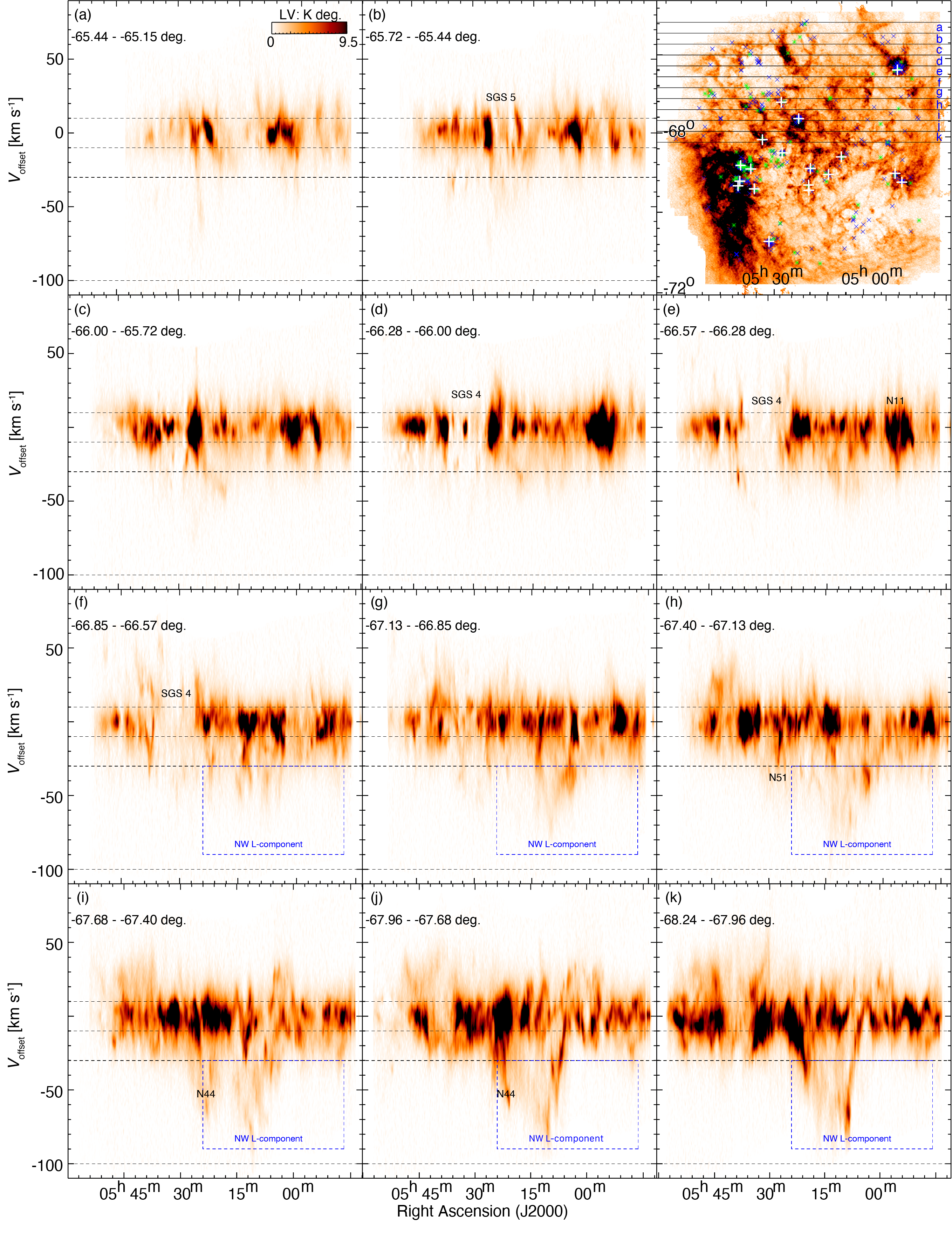}
\end{center}
\caption{The upper right panel shows total integrated intensity map of the LMC. {The green asterisks and blue crosses indicate WR stars and O-type stars (Bonanos et al. 2009)}. Horizontal lines indicate the integration ranges of Right Ascension--velocity diagrams in Dec. (a)--(k) Channel maps of Right Ascension--velocity diagrams over the whole H{\sc i}. The integration range is 0.27 deg. ($\sim$235.6 pc), and the integration range is shifted from north to south in 0.27 deg. step. The black dashed lines indicate the velocity ranges of the L-, I-, and D-components. The upper left number denotes the integration range in the upper right panel.}  
\label{fig21}
\end{figure*}%

\setcounter{figure}{17}
\begin{figure*}[htbp]
\begin{center}
\includegraphics[width=15cm]{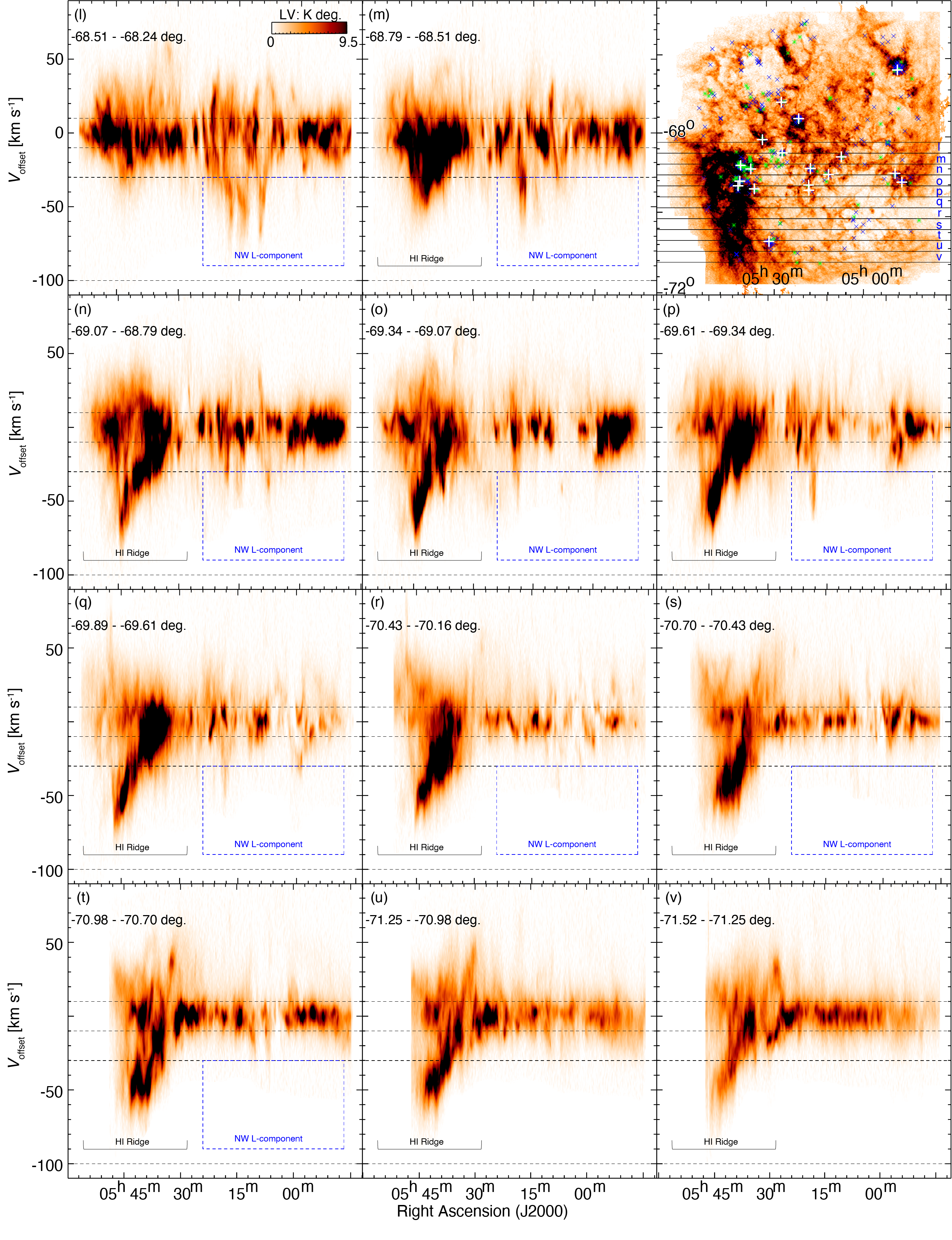}
\end{center}

\caption{Continued.}  
\label{fig21}
\end{figure*}%

{}

\end{document}